\documentclass[12pt]{article}
\usepackage{graphicx}
\usepackage{amsmath}
\usepackage{longtable}
\usepackage{color}
\usepackage{bbm}
\usepackage[...]{youngtab}
\usepackage{cancel}
\usepackage{multirow}

\relax
\usepackage{amssymb}

\definecolor{lred}{RGB}{180,10,10}
\definecolor{lblue}{RGB}{5,5,160}
\usepackage[colorlinks,,linktoc=section,linkcolor=lblue,citecolor=lblue,filecolor=lblue,urlcolor=lblue]{hyperref}

\newcommand{\bo}{{\bar o}}

\def\bo{{\raise.15ex\hbox{\large$\Box$}}}               

\def\face{{\raise.2ex\hbox{$\displaystyle \bigodot$}\mskip-2.2mu \llap {$\ddot
        \smile$}}}                                      

\def\leftrightarrowfill{$\mathsurround=0pt \mathord\leftarrow \mkern-6mu
        \cleaders\hbox{$\mkern-2mu \mathord- \mkern-2mu$}\hfill
        \mkern-6mu \mathord\rightarrow$}       
\def\dvec#1{\vbox{\ialign{##\crcr
        \leftrightarrowfill\crcr\noalign{\kern-1pt\nointerlineskip}
        $\hfil\displaystyle{#1}\hfil$\crcr}}}           

\def\beq{\begin{equation}}
\def\eeq{\end{equation}}

\def\beqx{\begin{equation*}}
\def\eeqx{\end{equation*}}

\def\beql{\begin{eqnarray}}
\def\eeql{\end{eqnarray}}
\def\beqa{\begin{eqnarray*}}
\def\eeqa{\end{eqnarray*}}
\def\NO{\nonumber}

\newcommand{\bea}{\begin{eqnarray}}
\newcommand{\eea}{\end{eqnarray}}

\def\[{\left [}
\def\]{\right ]}
\def\({\left (}
\def\){\right )}

   \def\eps{\epsilon}

\def\+{\oplus}

\begin{document}

\hbox{\hskip 12cm NIKHEF/2017-61  \hfil}
\hbox{\hskip 12cm IFF-FM-2017/07  \hfil}

\vskip .5in

\begin{center}
{\Large \bf Symmetry breaking by bi-fundamentals}

\vspace*{.4in}
{A.N. Schellekens}$^{a,b}$
\\
\vskip .2in
${ }^a$ {\em NIKHEF Theory Group, Kruislaan 409, \\
1098 SJ Amsterdam, The Netherlands} \\
\vskip .2in
${ }^b$ {\em Instituto de F\'\i sica Fundamental, CSIC, \\
Serrano 123, Madrid 28006, Spain} \\
%
%
%
%
\end{center}

\begin{center}
\vspace*{0.3in}
{\bf Abstract}
\end{center}
We derive all possible symmetry breaking patterns for all possible Higgs fields that can occur in intersecting brane models: bi-fundamentals and
rank-2 tensors. This is a field-theoretic problem that was already partially solved in 1973 by Ling-Fong Li \cite{Li:1973mq}.
 In that paper the solution was given for rank-2 tensors of orthogonal
and unitary group, and $U(N)\times U(M)$ and $O(N)\times O(M)$ bi-fundamentals. We extend this first of all to symplectic groups. 
When formulated correctly, this turns out to be straightforward generalization of the previous results from real and complex numbers to quaternions. 
The extension to mixed bi-fundamentals is more challenging and interesting. The scalar potential has up to six real parameters. Its minima or saddle points are
described by block-diagonal matrices built out of $K$ blocks of size $p\times q$. Here $p=q=1$ for the solutions of Ling-Fong Li, and the number of possibilities for
$p\times q$ is equal  to the number of real parameters in the potential, minus 1. The maximum block size is $p\times q=2\times 4$. Different blocks cannot be combined, and the true minimum occurs for one choice of basic block, and  for either $K=1$ or $K$ maximal, depending  on the parameter values.

\vskip .5truecm

\newpage

\tableofcontents

\definecolor{red}{RGB}{1,0,0}

\section{Introduction}

The purpose of this paper is to complete the classic work of Ling-Fong Li, \cite{Li:1973mq} (see also  \cite{Elias:1975yd} for corrections),  in which  symmetry breaking patterns caused
by Higgs fields in various representations of various groups are considered. Perhaps surprisingly, there does not exist a general formula or algorithm that
deals with this group-theoretical problem in full generality, quite unlike computing the tensor product for arbitrary representations, for example. 
In  \cite{Li:1973mq} some special cases were selected based on what seemed interesting at that time. With remarkable foresight, the author considered
representations that can occur as massless states in open string models, or more precisely intersection brane models: rank-2 tensors
and bi-fundamentals. However, the results of  \cite{Li:1973mq} do  not cover all possible brane configurations: symplectic groups were not discussed, and neither were
mixed group types: Unitary-Orthogonal, Unitary-Symplectic and Orthogonal-Symplectic. Here we will complete these results by working out the missing ones. 

In  field theory  any study of Higgs symmetry breaking is inevitable limited to an infinitesimal subset of the allowed Higgs representations. But 
in string theory those representations are limited to the ones that can be massless. 
In intersecting brane models, rank-2 tensors and bi-fundamentals   are the only representations that can occur. So one can actually solve the problem in full generality, if there is only a single Higgs field.

The concrete reason for embarking on this study  originates from attempts to generalize a previous paper \cite{Gato-Rivera:2014afa}. In that paper a surprisingly successful
attempt was made to derive the Standard Model from requirements on the complexity
of the low-energy physics it produces.  This is the opposite of trying to understand the Standard Model from some high energy symmetry principle, as one usually does.  
In a limited set of two-stack brane models, the $N$-family Standard Model turned out to stand out clearly as the case with the 
richest kind of ``atomic physics". This may be viewed as concretely defined proxy for an anthropic argument.
A crucial role was played by the assumption that a single Higgs field renders the entire spectrum non-chiral. 

When we attempted to extend this analysis to multi-stack models, we ran into a rather stubborn problem. The generalization required considering Higgs fields in any representation 
allowed by the brane configurations. Since branes (or open strings) can have unitary, orthogonal and symplectic groups (U,O and S for short) \cite{Marcus:1986cm}, and since 
bi-fundamentals can arise from any combination of branes, we needed the aforementioned generalization of \cite{Li:1973mq}. Although in some cases educated
guesses can be made based on group theory arguments, this is not satisfactory. 

While studying this problem it became clear that the general case is substantially more complicated than the U-U and O-O cases studied in \cite{Li:1973mq}. Instead
of two terms in the quartic potential for U-U, O-O and S-S, one gets three for U-O and U-S, and five for O-S, with up to 6 real coupling constants. Although initially this seemed intractable, we found that it can in  fact be solved
completely and exactly, and that it even simplifies the original analysis of  Ling-Fong Li. In fact, a rather beautiful solution emerges. 

Although the inspiration from this work came from string theory, all of it is in fact classical field theory. But there is a  remnant of the stringy origin, namely that fact that
we use the full unitary group and not the simple group $SU(N)$, because $U(N)$ is what comes naturally out of string theory.  One can get $SU(N)$ in open string theory in four dimensions if  the $U(1)$ phase symmetry is broken by axion mixing, but this is an entirely
separate and model-dependent issue which we do not consider here. 
Furthermore not only is the full gauge group of the theory a product of $U(N)$, $O(N)$ and $USp(N)$ factors, we find  that the same is true for the  stabilizer of the Higgs vacuum, the
broken subgroup. We find that $SU(N)$ groups occur only as part of a $U(N)$, or as $USp(N)$ for the special case $N=2$, because $USp(2)$ is isomorphic to $SU(2)$. This
suggests that the broken subgroups may have a simply string theory interpretation. Indeed, this hints at an interpretation in terms of a phenomenon known as ``brane recombination", discussed for example in \cite{Cremades:2002cs}.  This is an intriguing possibility, but we will not explore this further in the present work, since this lies beyond the scope of field theory.

This paper has two parts. In the first part we discuss the most interesting case, namely bi-fundamentals between different branes. The second part is about self-intersections, or in other words rank-2 
tensors. Here the only cases not yet considered in \cite{Li:1973mq} are (anti)-symmetric tensors of symplectic groups. The breaking patterns and energy considerations
turn out to be a natural extension of those for unitary and orthogonal groups. Besides symplectic rank-2 tensors the other novelty in this part is a greatly simplified
derivation, made possible by some fairly old, but not widely known  theorems on matrix (skew)-diagonalization.

For readers who just need the results and who are not interested in the methods and derivations,  two summary sections are provided, namely \ref{BiSum} for bi-fundamentals and \ref{TwoSum} for rank-2 tensors.

\section{Bi-fundamentals}

We consider a Higgs field in a bi-fundamental representation  of groups $G(N)\times H(M)$, where each group can be orthogonal, unitary or symplectic.  The global groups are $G(N)=O(N)$, $U(N)$ or $USp(N)$ and analogously for $H(M)$. The precise definition of $USp(N)$ is
given in the appendix. The symplectic groups only exists for even matrix dimensions, but to keep the notation universal we will use $N$ (or $M$) in all cases.

\subsection{The potential}

\subsubsection{Invariant contractions}\label{InvCon}

Consider a Higgs field  $\phi_{i\alpha}$ in a bi-fundamental representation  of groups $G(N)\times H(M)$,  
with $i=1,\ldots,M$ and $\alpha=1,\ldots,N$. 
We consider a renormalizable Higgs potential. This implies that it  can only have quadratic and quartic terms. Cubic terms are
not possible with bi-fundamentals. 
 
 Since all groups are embedded in a unitary group, one always obtains invariance under the left or right algebra if $\phi$ is combined with $\phi^*$, and the left and
right indices are contracted. This can be done in only one way for quadratic terms, and in two ways for quartic terms.

In addition, depending on the algebra one considers, it may be possible to obtain gauge invariant combinations by contracting indices of two $\phi$'s (or two $\phi^*$'s) with each other, provided an invariant contraction matrix exists. We will denote the contraction matrix for the
first index as $D_{ij}$, and the one for the second index as $C_{\alpha\beta}$. In a suitable basis, these matrices can be chosen as  either a Kronecker $\delta$  for orthogonal groups, or 
a skew-diagonal matrix with diagonal $i\sigma_2$ blocks for symplectic groups. 
The latter will be denoted $\Omega$.  One can of course define other bases, and in particular for symplectic groups $USp(2K)$ there are two that are often used. One is the matrix $\Omega$, which can be written as $\mathbbm{1}_K \otimes (i\sigma_2)$, and the other canonical choice is
$(i\sigma_2) \otimes \mathbbm{1}_K$; this is   the matrix $h_A$ shown in Eqn. (\ref{hAdef}). 
Of course no results depend on that choice, but the form $\Omega$ is convenient in some computations, and also makes a formulation in terms of quaternions possible in certain cases.

We define 
\beqa C^T&=&\epsilon_C C \\
D^T&=&\epsilon_D D
\eeqa
where $\epsilon_C$ and $\epsilon_D$ are signs; they are $+1$ if $C$ or $D$ are Kronecker $\delta$'s, and $-1$ if $C$ or $D$ or equal to a matrix $\Omega$ (or $h_A$).

\subsubsection{Mass terms and reality conditions}

If both $C$ and $D$ exist, we can write down two distinct mass terms
\beqa
& \phi_{i\alpha}\phi_{i\alpha}^* = {\rm Tr}\ \phi \phi^{\dagger} \\
& \phi_{i\alpha}\phi_{j\beta}D_{ij}C_{\alpha\beta} = {\rm Tr} \phi C \phi^T D
\eeqa
The second one vanishes unless $\epsilon_C=\epsilon_D$, so it exists only if both groups are orthogonal or both groups are symplectic. The existence of an additional mass term indicates that there are two separate fields rather than just one. We can eliminate one
of these components by imposing a reality condition
\beq\label{RealityCondition}
\phi^*=D^T \phi C
\eeq
With the canonical choices for $C$ and $D$ given above, these reality conditions simply imply that $\phi$ is real or quaternionic for orthogonal or symplectic groups respectively (see the appendix for more details).
In the latter case this means that $\phi$ can be written in terms of quaternions, defined on the $2\times 2$ blocks of the matrices $\Omega_{ij}$
and $\Omega_{\alpha\beta}$.
Both kinds of reality conditions 
imply a reduction by a factor of two in the number of degrees of freedom. A general complex field may be written as a real part plus an imaginary field; the quaternionic 
analog is that its a $2\times 2$ block can be written as a $q+iq'$, where $q$ and $q'$ are quaternions.

\subsubsection{Quartic terms}

If both $C$ and $D$ exist, the most general invariant potential has the form
\begin{eqnarray}\label{ThePotential}
V=& -&\mu^2 {\rm Tr} \phi\phi^{\dagger}   +  \tfrac12  \lambda_1 ({\rm Tr} \phi\phi^{\dagger})^2+  \tfrac12  \lambda_2 {\rm Tr} (\phi\phi^{\dagger})^2+ \tfrac12  \lambda_3 {\rm Tr}(\phi C \phi^T)(\phi C \phi^T)^{\dagger}\NO\\&+& 
 \tfrac12  \lambda_4 {\rm Tr}(\phi^T D\phi)(\phi^T D \phi)^{\dagger}+  \tfrac14   \left [\lambda_5{\rm Tr} (\phi C \phi^T D)^2 + {\rm h.c.}\right]
\end{eqnarray}
The normalization of this potential differs from the one used in \cite{Li:1973mq}. We have chosen a normalization so that for complex fields all numerical factors in the equation of motion are equal to 1. For
this reason our potential is larger by an overall factor 2 with respect to \cite{Li:1973mq}. Since we are only interested in the extrema, this is irrelevant. For future purposes we write the potential as
\begin{equation}\label{PotentialTwo} 
V=-\mu^2  \phi_{i\alpha} \phi^*_{i\alpha}+  \tfrac12  \lambda_1  V_1
+  \tfrac12  \lambda_2  V_2
+ 
 \tfrac12 \lambda_3  V_3   + \tfrac12  \lambda_4 V_4 
+    \tfrac14  \lambda_5  V_5 + \tfrac14  \lambda_5^*V_5^*
\end{equation}
Written in terms of components the potential is
\begin{eqnarray}\label{Potential} \NO
V=& -&\mu^2  \phi_{i\alpha} \phi^*_{i\alpha}+  \tfrac12  \lambda_1  \phi_{i\alpha} \phi^*_{i\alpha}\phi_{j\beta} \phi^*_{j\beta}
+  \tfrac12  \lambda_2  \phi_{i\alpha} \phi^*_{i\beta}\phi_{j\beta} \phi^*_{j\alpha}\\ \NO
&+& 
 \tfrac12 \epsilon_C \lambda_3  \phi_{i\alpha} C_{\alpha\beta}\phi_{j\beta}   \phi^*_{j\gamma} C_{\gamma\delta}\phi^*_{i\delta}   + \tfrac12 \epsilon_D \lambda_4 \phi_{i\alpha} D_{ij}\phi_{j\beta}\phi^*_{k\beta}D_{kl}\phi^*_{l\alpha} 
 \\&+&  \epsilon_D \tfrac14  \lambda_5   \phi_{j\alpha} C_{\alpha\beta}\phi_{k\beta}D_{kl}   \phi_{l\gamma} C_{\gamma\delta}\phi_{i\delta} D_{ij} + \tfrac14  \lambda_5^*\phi_{j\alpha}^* C_{\alpha\beta}\phi^*_{k\beta} D_{kl}  \phi^*_{l\gamma} C_{\gamma\delta}\phi^*_{i\delta}D_{ij}
\end{eqnarray}
Note that the $\lambda_1$ term differs from all other four-point couplings because it has two disjoint index loops rather than one. The equations
of
motion break one loop into a string. Hence the $\lambda_1$ term contributes a factor  with a closed index loop to the equations of motion. Therefore it
contributes to the equations of motion through a trace over all non-zero elements of $\phi$. This explains why the $\lambda_1$ dependence is 
different from all other coupling dependencies.

In its most general form the potential has four real parameters, $\lambda_1\ldots\lambda_4$, and one complex one, $\lambda_5$. The most general form applies only to cases with both a $C$ and  a $D$ matrix, namely $O(N)\times O(M)$, $USp(2N)\times USp(2M)$ and $O(N)\times USp(2M)$. However, in the first two of these cases
we  have to impose a reality condition. This implies that we can express  all $\phi^*$ in terms of $\phi$. 
Then there are only two distinct quartic terms possible, namely $V_1$, written as $(\phi\phi)^2$, and $V_5$. All other terms can be expressed in terms of products of
four fields $\phi$, and since they have just one index loop they must all be related to $V_5$.
Note that if a  reality condition is imposed $V_5$ and $V_5^*$
 are separately real, and $\lambda_5$ can be chosen real without loss of generality.  Since $V_2, V_3$ and $V_4$ are all related to $V_5$,  we
may select one of them to our convenience.
So we make a choice that is universally valid, and keep just the $\mu^2, \lambda_1$ and $\lambda_2$ term for  $O(N)\times O(M)$ and $USp(2N)\times USp(2M)$. All terms in the potential are distinct only if the symmetry group  is $O(N)\times USp(2M)$

In the table below we list all potential terms that can occur. For future purposes we have chosen a certain order of the two group types in the 
last three cases.
 \begin{table}[h]
\begin{center}
\begin{tabular}{|c|c|c|c|c|} \hline
Groups & $D$ & $C$ & reality condition & coupling constants \\ \hline
$U(N)\times U(M)$ & none & none & none & $\lambda_1, \lambda_2$ \\
$O(N)\times O(M)$ & $\delta$ & $\delta$ & $\phi=\phi^*$ & $\lambda_1, \lambda_2$ \\
$USp(2N)\times USp(2M)$ & $\Omega$ & $\Omega$ & $\phi^*=\Omega^T\phi\Omega$ & $\lambda_1, \lambda_2$ \\
$O(N)\times U(M)$ & $\delta$ & none & none & $\lambda_1, \lambda_2,\lambda_4$ \\
$U(N)\times USp(2M)$ & none & $\Omega$ & none & $\lambda_1, \lambda_2,\lambda_3$ \\
$O(N)\times USp(2M)$ & $\delta$ & $\Omega$ & none & $\lambda_1, \lambda_2,\lambda_3, \lambda_4, \lambda_5$ \\
 \hline
\end{tabular}
\caption{Potential terms\label{CouplingTable}}
\end{center}
\end{table}

\subsubsection{The vacuum energy}

Before getting into the details of solving the equations of motion, we derive here a useful general formula for the vacuum energy
at the extremal points of potentials with only quadratic and quartic terms. Consider potentials of the form
\begin{eqnarray*} 
V=-\mu^2\phi_x\phi_x^* &+& \tfrac12  \lambda_1  (\phi_x\phi_x^*)^2  +  \tfrac12 \sum_i \lambda_i T^i_{xyvw}\phi_x\phi_y\phi_v^*\phi_w^*\\
&+& \tfrac14 \sum_i \rho_i P^i_{xyvw}\phi_x\phi_y\phi_v\phi_w+ \tfrac14 \sum_i \rho^*_i P^i_{xyvw}\phi^*_x\phi^*_y\phi^*_v\phi^*_w
\end{eqnarray*} 
The potential given above is of this form, with $x$ interpreted as the index pair $i\alpha$. 
Here we assume that $P^i$ and $T^i$ are real, as indeed they are in the case of interest.  All indices $x,y,v,w$ are implicitly summed. 
Note that $T^i$ is symmetric in $x,y$ and $v,w$, and that
$P^i$ is symmetric in all four indices $xyvw$. Furthermore we assume that $T^i$ satisfies $T^i_{xyvw}=T^i_{vwxy}$, so that
all terms involving $T^i$ are real. 
The equations of motion, derived by differentiating with respect to $\phi_z^*$ are 
\begin{eqnarray*} 
\frac{\partial V}{\partial \phi_z^*}=0=-\mu^2\phi_z &+&  \lambda_1  \phi_z(\phi_x\phi_x^*)  +\  \sum_i \lambda_i T^i_{xyvz}\phi_x\phi_y\phi_v^*\\
&+&   \sum_i \eta^*_i P^i_{xyvz}\phi^*_x\phi^*_y\phi^*_v
\end{eqnarray*} 
If we multiply this with $\phi_z^*$ and sum over $z$ we get
\begin{equation} \label{MusqPot}
0=-\mu^2\phi_z \phi_z^*  +  \lambda_1 (\phi_x\phi_x^*)^2  +\  \sum_i \lambda_i T^i_{xyvz}\phi_x\phi_y\phi_v^*\phi_z^*
+   \sum_i \eta^*_i P^i_{xyvz}\phi^*_x\phi^*_y\phi^*_v\phi_z^*
\end{equation} 
The first three terms on the lefthand-side are real. Hence the equations of motion imply that 
\begin{equation} \label{Reality}
{\rm Im} \sum_i \eta^*_i P^i_{xyvz}\phi^*_x\phi^*_y\phi^*_v\phi_z^* = 0
\end{equation} 
This implies that we may write
\begin{equation*} 
  \sum_i \eta^*_i P^i_{xyvz}\phi^*_x\phi^*_y\phi^*_v\phi_z^*= \tfrac12 \sum_i \eta^*_i P^i_{xyvz}\phi^*_x\phi^*_y\phi^*_v\phi_z^*+ \tfrac12 \sum_i \eta_i P^i_{xyvz}\phi_x\phi_y\phi_v\phi_z
\end{equation*} 
Substituting this into Eqn. (\ref{MusqPot}), we see that we can express all quartic potential contributions in terms of the mass terms, and that the value of $V$ at an extremum is always given by 
\beq\label{VacEn}
 V_{\rm EOM}=-\tfrac12 \mu^2\phi_x\phi_x^*  \ ,
 \eeq
 with implicit summation over $x$, as before.
This conclusion does not hold if there are cubic terms in the potential.

\subsection{Symmetric bi-fundamentals}\label{SymBiFund}

In this section we  consider the case of a bi-fundamental Higgs system with symmetry group $U(N)\times U(M)$, $O(N)\times O(M)$ and $USp(N)\times USp(M)$. 
We call these {\em symmetric} bi-fundamentals because the left and right group are of the same type. However, $N$ and $M$ may be different.
Note that the 
two groups act independently. We will make use of the fact that the field $\phi$ is complex, real and quaternionic in these three cases respectively.
This allows a simultaneous derivation of the result in all three cases. As explained above, we can use the reality or quaternionic constraint to show that only the
$\lambda_1$ and $\lambda_2$ terms in the potential contribute.
The equations of motion are
\beqx
\mu^2 \phi_{i\alpha}  =   \lambda_1  \phi_{i\alpha}\left(\phi_{j\beta} \phi^*_{j\beta}\right)+  \lambda_2 \phi_{i\beta}\phi_{j\alpha}\phi^*_{j\beta}
\eeqx
This is obtained from the $\phi_{i\alpha}^*$ variation. We treat the field here as complex in all cases. In the quaternionic cases the fields belong to
$2\times 2$ blocks, but there is no need for indicating that. 
Strictly speaking the $\phi_{i\alpha}^*$ variation also acts on $\phi$ if there is a reality condition. But this just gives rise to an extra factor 2 in all 
contributing terms, and hence one gets exactly the same equation. 

The easiest way to solve the problem is to make use of a singular value decomposition of $\phi$. This means that $\phi$ can be written as
\beqx
\phi=U R V
\eeqx
where $U \in O(N), U(N)$ or $USp(N)$ and $V\in O(M), U(M)$ or $USp(M)$ in the three cases respectively. The matrix $R$ is diagonal, real and non-negative. Note that
negative diagonal elements can be made positive by a suitable one-sided $U(N)$ or $O(N)$ transformation. In the symplectic case $R$ must be diagonal in terms of quaternions , which means that it consists of $2\times 2$ blocks $r \mathbbm{1}$, with $r$ real and positive. Also in this case a negative sign can be flipped by a one-sided 
symplectic transformation (namely $-\mathbbm{1} \in SU(2) \subset USp(N)$).  
The singular value decomposition for quaternions has been
derived in \cite{Zhang}. 
Note that a singular value decomposition works even if $\phi$ is not a square matrix.

Although singular value decomposition are less widely known than matrix
diagonalizations, they are used in the Standard Model of particle physics, namely for the ``diagonalization" of quark mass matrices. Since the latter are square matrices, 
one usually solves this problem by means of polar decompositions, which can be viewed as a special case of singular value decompositions. In \cite{Li:1973mq} no use was made of 
singular value decompositions. Instead the Hermitean quantity $\phi\phi^{\dagger}$ was used, which can be diagonalized in the traditional way.  But  an extra step is needed to relate the diagonal form of $\phi\phi^{\dagger}$ to $\phi$ itself.

We can now arrive at the final answer in just a few steps. By a left and right gauge transformation one can remove $U$ and $V$, and hence one may  
replace $\phi$ by its eigenvalues: $\phi_{i\alpha}=r_i \delta_{i\alpha}, r_i \in \mathbbm{R}; r_i \geq 0$. 
Substituting this into the equations of motion we get
\beqx
\mu^2 r_i \delta_{i\alpha}  =   \lambda_1  r_i \delta_{i\alpha} \sum_j r_j^2 +  \lambda_2 r_i^3 \delta_{i\alpha}
\eeqx
For those values of $i$ with $r_i\not=0$ this implies
\beqx
\mu^2   =   \lambda_1   \sum_j r_j^2 +  \lambda_2 r_i^2 
\eeqx
which implies that all non-vanishing $r_i$ must be identical: $r_i=r$. Suppose there are $K$ non-vanishing ones. Then 
\beqx
\mu^2   =   \lambda_1   Kr^2 +  \lambda_2^2 r^2
\eeqx
which means that (since $r \geq 0$)
\beqx
r=\sqrt{\frac{\mu^2}{K\lambda_1+\lambda_2}}
\eeqx
The energy of these solutions is, from (\ref{VacEn}) or directly from the potential:
\beqx
E=-\tfrac12{\frac{ \mu^4 K}{K\lambda_1+\lambda_2}}
\eeqx

\subsection{Asymmetric bi-fundamentals}

\subsubsection{Why a new strategy is needed}

All methods used so far for the symmetric case, diagonalization of $\phi$ or $\phi\phi^{\dagger}$ or singular value decompositions, fail
in all asymmetric cases, U-O, U-S and O-S. Note first of all that $\phi$ is a general complex matrix in all these cases. No reality or quaternionic constraints 
can be imposed. Therefore $\phi$ is not diagonalizable by means of the available gauge symmetries. Quantities like $\phi\phi^{\dagger}$ are diagonalizable 
in some cases, but the potential has additional terms, and cannot be expressed in terms of just one of the matrices $\phi\phi^{\dagger}$, $\phi^{\dagger}\phi$, $\phi\phi^{T}$
or $\phi^T\phi$. If there is more than one, they would have to be diagonalized simultaneously. Furthermore for the diagonalization of some of these quantities unitary
transformations are required, while only orthogonal or symplectic ones are available.

None of the known theorems concerning singular value decompositions apply to U-O, U-S or O-S. In other words, one may of course ``diagonalize" any complex matrix $\phi$ by means of unitary
matrices, but it will not be possible to gauge these transformation matrices away. One may speculate about the possibility that
there exists some generalization of singular vector decompositions that allows us to write $\phi$ in a simpler form. While some version of that statement may be correct, some concrete guesses
can be ruled out by counting parameters. For example, compare $U(N)\times U(N)$ and $O(N) \times USp(N)$ (we focus on square matrices here). In both cases the field $\phi$ is complex and has
$2N^2$ real parameters, precisely the same as the number of parameters of $U(N)\times U(N)$.  The difference between these number determines the minimal number of parameters that remains after
gauge fixing; the actual number can be larger because of degeneracies in the action of the gauge symmetries.   Indeed, in this case the minimal number of parameters is 0, but the actual number is $N$
because common phases in the left- and right unitary groups have the same effect. For $O(N)\times O(N)$ the minimal number is $N$, which is also the actual number, and 
for $USp(N)\times USp(N)$ with a quaternionic condition the minimal number is $-N$, and the actual number $\tfrac12 N$.

But for $O(N) \times USp(N)$ the minimal number of remaining parameters is $$2N^2-\tfrac12 N(N-1) - \tfrac12 N(N+1)=N^2\ . $$ This immediately ruins any chance for a diagonal form, and even for some
more general block-diagonal form. If the putative final result is a block-diagonal matrix with $(N/p)$ complex $p\times p$ blocks, the total number of parameters is  $2Np$. Only for $p=N/2$ one can just
saturate the bound, but then the matrix just decomposes into two general complex $N/2 \times N/2$ matrices. Even if this were possible, it does not look like a useful result. So we need an entirely
different strategy.

 \subsubsection{Equations of motion}
 We have just seen that the procedure of first bringing the fields in the simplest form using gauge rotations, and only then applying the equations of motion,
 fails. Our strategy will therefore be to intertwine these two tools in several steps: simplify by gauge rotations, then apply the equations of motion, then simplify further by another
 gauge rotation, and then use the equations of motion once more.
  
 We consider the potential with all five quartic terms, but one has to keep in mind that in some cases some of the parameters may vanish. 
The equations of motion, obtained by varying the potential with respect to $\phi_{i\alpha}^*$ are
\def\firstindex{i}
\begin{eqnarray}\label{EOM}
\mu^2 \phi_{\firstindex\alpha}  &=&   \lambda_1  \phi_{\firstindex\alpha}\left(\phi_{j\beta} \phi^*_{j\beta}\right) +  \lambda_2 \phi_{\firstindex\beta}\phi_{j\alpha}\phi^*_{j\beta}
+\epsilon_C   \lambda_3 \phi_{\firstindex\delta} C_{\delta\beta}\phi_{j\beta}\phi_{j\gamma}^* C_{\gamma\alpha}\NO \\
&+& \epsilon_D  \lambda_4 \phi_{m\alpha}D_{mj} \phi_{j\beta}\phi_{k\beta}^*D_{ki}+  \lambda_5^* C_{\alpha\beta}\phi^*_{j\beta} D_{jk}  \phi^*_{k\gamma} C_{\gamma\delta}\phi^*_{m\delta}D_{mi}
\end{eqnarray}
We may distinguish two kinds of equations of motion: those for $\phi_{i\alpha}=0$ and those for $\phi_{i\alpha}\not=0$. 
In equations of the former kind, both the $\mu^2$ term and the $\lambda_1$ term drop out, and one is left with the last four terms. We will call two kinds of equations ``homogeneous" and
the ``inhomogeneous" respectively because in the first kind all terms are cubic. This is a slight abuse of the terminology commonly used for linear equations.

We will not consider solutions that only work for special values of $\lambda_i$ or special relations among the parameters $\lambda_i$. Such relations are not renormalization group invariant unless
the potential has some additional symmetry. This requirement rules out cancellations among the four cubic terms with coefficients $\lambda_2 \ldots \lambda_5$, the four terms of a homogeneous equation. 
Each must vanish separately, and
hence for each homogeneous equation we get up to four equations, one for each coupling constant. 

Most of the information in the inhomogeneous equations can be dealt with in the 
same way: one can derive a second class of homogeneous equations from them.  
Consider two non-vanishing elements $\phi_{i\alpha}$ and  $\phi_{k\gamma}$. From two inhomogeneous equations we can obtain a homogeneous one by multiplying the equations for $\phi_{i\alpha}$
by $\phi_{k\gamma}$ and vice-versa, and subtracting the two. Then the $\mu^2$ and $\lambda_1$ terms drop out.
The resulting difference equation must be satisfied for generic values of $\lambda_p$, and hence it splits into
four separate equations, labelled by the index $p$ of the coupling constants $\lambda_p$.

This argument cannot be applied to the inhomogeneous equations because they have parameters of different dimensions. If all terms in the equation have the same tensor structure they
can be made to cancel by changing the overall scale of the field. Hence such cancellations do not depend on special, fixed relations between $\mu^2$ and the coupling constants.

In order to make the metric $C$ and $D$ explicit we will assume that the symplectic groups act only on the column basis (labelled by $\alpha,\beta,\ldots$), 
and orthogonal ones only on the row basis $(i,j,\ldots$). This allows us to consider the four cases $U(N)\times U(M)$, $O(N)\times U(M)$, $U(N)\times USp(M)$ and $O(N)\times USp(M)$.
Although $U(N)\times U(M)$  has already been solved, we include it for illustrative purposes. Now we can replace $D$ by a Kronecker $\delta$ and 
$C$ by an anti-symmetric matrix $\Omega$. But since $\Omega$ pairs indices, we will absorb it in most cases in the fields by defining 
\beqx
\phi_{i\tilde\alpha}=\Omega_{\alpha\beta}\phi_{i\beta}
\eeqx
so that the indices $\alpha$ and $\tilde\alpha$ form a symplectic pair.
The remaining two types, $O(N)\times O(M)$ and
$USp(N)\times USp(M)$  cannot be treated in this way. Nevertheless these cases can also be solved by the following method, by applying it to real numbers or quaternions. 
But because these cases have already been solved, we will not discuss this.

For the special case where $G(N)$ is either $U(N)$ or $O(N)$ and $H(M)$ is either $U(M)$ or  $USp(M)$ the equations of motion for $\phi_{i\alpha}$ read
\begin{eqnarray}\label{Homogen}
\mu^2 \phi_{\firstindex\alpha}  =   \lambda_1  \phi_{\firstindex\alpha}\left(\phi_{j\beta} \phi^*_{j\beta}\right) &+&  \lambda_2 \phi_{\firstindex\beta}\phi_{j\alpha}\phi^*_{j\beta}
+  \lambda_3 \phi_{\firstindex\beta} \phi_{j\tilde\beta}\phi_{j\tilde\alpha}^* \NO \\
&+&   \lambda_4 \phi_{j\alpha} \phi_{j\beta}\phi_{i\beta}^*+  \lambda_5^* \phi^*_{j\tilde\alpha}   \phi^*_{j\beta} \phi^*_{i\tilde\beta}
\end{eqnarray}
The homogeneous equations are simply that the four terms with coefficients $\lambda_p$, $p=2\ldots 5$ must vanish if $\phi_{i\alpha}=0$ (note that the $\lambda_1$ term also vanishes in that case).
The second class of homogeneous equations mentioned above, the weighted difference of two inhomogeneous equations for non-zero fields $\phi_{i\alpha}$ and $\phi_{k\gamma}$, yields
\beqa
G\times H: \quad\phi_{\firstindex\beta}\phi_{j\alpha}\phi^*_{j\beta}\phi_{k\gamma}&=&\phi_{k\beta}\phi_{j\gamma}\phi^*_{j\beta}\phi_{i\alpha}\\
G\times S: \quad\phi_{\firstindex\beta} \phi_{j\tilde\beta}\phi_{j\tilde\alpha}^*\phi_{k\gamma}&=&\phi_{k\beta} \phi_{j\tilde\beta}\phi_{j\tilde\gamma}^*\phi_{i\alpha}\\
O\times H: \quad \phi_{j\alpha} \phi_{j\beta}\phi_{i\beta}^*\phi_{k\gamma}&=& \phi_{j\gamma} \phi_{j\beta}\phi_{k\beta}^*\phi_{i\alpha}\\
O\times S: \quad  \phi^*_{j\tilde\alpha}   \phi^*_{j\beta} \phi^*_{i\tilde\beta}\phi_{k\gamma}&=&  \phi^*_{j\tilde\gamma}   \phi^*_{j\beta} \phi^*_{k\tilde\beta}\phi_{i\alpha}
\eeqa
Not all these equations are available in all cases; this depends on $p$ as indicated in table \ref{CouplingTable}. We have 
indicated this here by $G=U$ or $O$, $H=U$ or $USp$, and $S$ is a used as a short-hand for $USp$.
These  relations are implicitly summed over $\beta$ and $j$. 

\subsubsection{Equations for pivot elements}\label{SpecialElt}

It will turn out  to be sufficient to study these equations for special cases where the row and column of a certain element $\phi_{ij}\not=0$ consists mostly of zeroes. Consider first 
 the special case where a row $i$ contains only one element, labelled by column index $\delta(i)$: 
 \beqx
 \phi_{i\delta(i)}\not=0; \quad 
\phi_{i\alpha}=0 \ \hbox{for all}\  \alpha\not=\delta(i), 
\eeqx
We call such an element $\phi_{i\delta(i)}$ a {\em pivot element}.

There is always at least one such element, because one can always bring one row into that form using either
 unitary or symplectic column transformations (note that in our setup orthogonal transformations do not  act on the column indices).  
 This works somewhat differently in the unitary and symplectic case, and the details are explained in the appendix. 
Obviously one can do this for just one row or column at a time. We may also use the row and column symmetries to
set $i=\delta(i)=1$, but we leave the notation general for now.

The existence of such a row implies  that there are $N-1$ homogeneous equations due to $\phi_{i\alpha}=0$. 
Furthermore, since there is only one non-zero element, the sum over $\beta$ collapses to a single term. 
Each of the four homogeneous equations can be divided by $\phi_{i\delta(i)}\not=0$. Then we get
\begin{eqnarray}
G\times H: \quad\phi_{j\alpha}\phi^*_{j\delta(i)}&=&0\label{OrthoOne}\\
G\times S: \quad \phi_{j\tilde\delta(i)}\phi_{j\tilde\alpha}^*&=&0\label{OrthoTwo}\\
O\times H: \quad \phi_{j\alpha} \phi_{j\delta(i)}&=&0\label{OrthoThree}\\
O\times S: \quad  \phi_{j\tilde\alpha}   \phi_{j\tilde\delta(i)}&=&0\label{OrthoFour}
\end{eqnarray}
This implies that in general, every column must be complex orthogonal ($\vec v\cdot\vec w^*=0$) to the column with the pivot element. In the symplectic case, every column must in addition by complex orthogonal to the column 
$\tilde \delta$ paired with the pivot element column. Furthermore the  column $\tilde \delta$ must be complex orthogonal to column $\delta$. In the $O(N)$ case the same statements must hold for both complex and
real orthogonality ($\vec v\cdot\vec w=0$).

Now consider inhomogeneous difference equations for pivot elements.  This requires 
 a second row $k$ with a pivot element in column $\delta(k)$. Having two such rows cannot be arranged purely by gauge rotations, but we will need this
 later in an intermediate step. 
 Assuming two rows with pivot elements $\phi_{i\delta(i)}$ and $\phi_{k\delta(k)}$ we get in the four cases respectively (with implicit sums over $j$)
\begin{eqnarray}
G\times H: \quad\  \phantom{ \phi_{j\delta(i)} \phi_{j\delta(i)}}\phi_{j\delta(i)}\phi^*_{j\delta(i)}&=&\phi_{j\delta(k)}\phi^*_{j\delta(k)}\label{InhomOne}\\
G\times S: \quad\  \phantom{ \phi_{j\delta(i)} \phi_{j\delta(i)}}  \phi_{j\tilde\delta(i)}\phi_{j\tilde\delta(i)}^*&=& \phi_{j\tilde\delta(k)}\phi_{j\tilde\delta(k)}^*\label{InhomTwo}\\
O\times H: \quad \phi_{j\delta(i)} \phi_{j\delta(i)}\phi_{i\delta(i)}^*\phi_{k\delta(k)}&=& \phi_{j\delta(k)} \phi_{j\delta(k)}\phi_{k\delta(k)}^*\phi_{i\delta(i)}\label{InhomThree}\\
O\times H: \quad  \phi^*_{j\tilde\delta(i)}   \phi^*_{j\tilde\delta(i)} \phi^*_{i\delta(i)}\phi_{k\delta(k)}&=&  \phi^*_{j\tilde\delta(k)}   \phi^*_{j\tilde\delta(k)} \phi^*_{k\delta(k)}\phi_{i\delta(i)}\label{InhomFour}
\end{eqnarray}
In the first two equations we have divided by the non-vanishing element $\phi_{i\delta(i)}$ and $\phi_{k\delta(k)}$. In the last two this is not possible, because these
factors appear as conjugates on the left- and right-hand side.  The first two equations imply that any two columns containing at least one pivot element must have the same norm. Furthermore, in the
symplectic case, their symplectic conjugate columns $\tilde \delta(i)$ and $\tilde \delta(k)$ must have the same norms as well. Note that this is true even if the columns $\tilde \delta(i)$ and $\tilde \delta(k)$  do not contain a 
pivot element themselves. If one of the symplectic conjugate columns $\tilde \delta(i)$ or $\tilde \delta(k)$ contains a pivot element, then all four columns $\delta(i)$, $\tilde \delta(k)$, $\tilde \delta(i)$ and $\tilde \delta(k)$
must have the same norm. 

A special case of interest is that of two or more pivot elements  appearing in the {\em same} column. Then the first two equations are trivially satisfied, but now the third and four equations add new information.
Suppose  we have at least two row labels $i$ and $k$ with $\delta(i)=\delta(k)\equiv\delta$. Since all other elements on these rows
vanish we can make unitary or orthogonal rotations on these rows, which allow us to bring the $\delta$-column vector $(\phi_{i\delta},\phi_{k\delta},\ldots)$ into a special form. The dots indicate any 
additional pivot elements
on column $\delta$. If $G=U$ we can rotate the column so that only $\phi_{i\delta}\not=0$.  This case is of no further interest, since now we have only a single pivot element.
If $G=O$ we have orthogonal transformations acting on complex vectors.
Then we can rotate the $\delta$-column to the form  $\phi_{i\delta}=x$ and $\phi_{k\delta}=r$. 
If $x=x^*$ we can rotate $r$ to zero,  so that we have only a single pivot element, and nothing new can be learned.
So assume that $x\not=x^*$ and $r\not=0$. Then
 we find from the last two equations, after dividing by $(x-x^*)r$ 
\beqa
\sum_j \phi_{j\delta} \phi_{j\delta} &=&0 \quad \hbox{for} \   O(N)\times U(N)\ \hbox{or}\ O(N)\times USp(N)\\
\sum_j \phi_{j\tilde\delta} \phi_{j\tilde\delta}&=&0\quad \hbox{for} \  O(N)\times USp(N)
\eeqa
These equations will be especially useful if a column contains exactly two pivot elements, because then it 
implies that these elements must differ by a factor $i$.

\subsubsection{The inhomogeneous equation}\label{InhomEq}

Now we turn to the inhomogeneous equations. They have the form (\ref{Homogen}). For a given solution, we may write all non-vanishing elements as 
\beqx
\phi_{i\alpha}= r\mu \chi_{i\alpha}
\eeqx
The functions $\chi_{i\alpha}$ are dimensionless, and can be given a standard 
normalization by setting one of them to 1. To define $r$ we have to find some canonical definition of one 
special non-zero element $\phi_{i\alpha}$. This can be done as follows. First we work out the row norms 
$n_i=\sum_{\alpha} \phi_{i\alpha}\phi^*_{i\alpha}$. These are invariant under all column basis transformations
$H(M)$, because all column transformations are either $U(M)$ or a subgroup of $U(M)$. We consider arbitrary 
$G(N)$ transformations of the rows $n_i$ to maximize the largest norm in this set. The row with the largest possible norm
is then $G(N)$-transformed to row 1.
This is possible for any choice of $G(N)$. Using a $H(M)$ transformation we now rotate row 1
so that only $\phi_{11}$ is non-zero, real and positive, as explained above. Now we define $r$ in such a way that $\chi_{11}=1$. Then we have obtained a basis so that
\beqx
\chi_{11}=1; \ \ \ \chi_{1\alpha}=0 \ \hbox{for all}\ \alpha \geq 2
\eeqx
This procedure implies that $|\chi_{i\alpha} | \leq 1$ for all 
$i$ and $\alpha$.  In terms of this parametrization the inhomogeneous equation now becomes
\beq\label{HomogenTwo}
 1  =   r^2 \left[\lambda_1   P  +  \lambda_2 \chi_{j1}\chi^*_{j1}
+  \lambda_3  \chi_{j2}\chi_{j2}^* 
+   \lambda_4 \chi_{j1} \chi_{j1}+  \lambda_5^* \chi^*_{j2}   \chi^*_{j2} \right]
\eeq
where $P=\sum \chi_{i\alpha}\chi^*_{i\alpha}$. Note that by construction $\phi_{11}$ is real, and by definition $\chi_{11}=1$. Furthermore $\mu$ is real, so
$r$ must be real as well. But this not manifest in this equation: the last two terms are not manifestly real. However, clearly their sum must be real,
and since we do not allow solutions that require special relations between parameter values, this must imply that they are separately real.
We define
\begin{eqnarray}\label{RhoDef}
 \rho_2 &=&   \chi_{j1}\chi^*_{j1}\NO \\
\rho_3 &=&   \chi_{j2}\chi_{j2}^* \NO \\
\rho_4 &=&   \chi_{j1} \chi_{j1} \NO\\
\rho_5 &=&  \chi^*_{j2}   \chi^*_{j2} 
\end{eqnarray}
Then we get
\beq\label{rEquation}
r=\sqrt{\frac{1}{P \lambda_1 + Q}}
\eeq
where 
\begin{eqnarray}\label{Qdef}
P&=&\sum \chi_{i\alpha}\chi^*_{i\alpha}\NO \\
Q&=&\sum_{i=2}^4  \lambda_i\rho_i + \lambda_5^*\rho_5 
\end{eqnarray}

In the special basis we have chosen, $\chi_{11}=1$, and $\chi_{1\beta}=0$ for $\beta > 1$. This implies that we can write out the sum over $\beta$
\beqx
\rho_2=\chi_{11} \sum_{j=1}^N \chi_{j1}\chi_{j1}^*=1+\sum_{j=2}^N \chi_{j1}\chi_{j1}^* \geq 1
\eeqx
Using row transformations acting on the last $N-1$ rows we can always bring the first column in a simpler form. If $G(N)=U(N)$, we can rotate all 
elements except $\chi_{21}$ to zero values. If $G(N)=O(N)$ we can bring $\chi_{21}$ to a general complex value, and $\chi_{31}$ to a positive real value. 
Since row 1 has norm 1, and since the norms were maximized using $G(N)$, we have  $|\chi_{21}| \leq 1$ and $|\chi_{31}|\leq 1$ .
Hence $\rho_2 \leq 3$. In practice the maximum value attained by $\rho_2$ will turn out to be 2.

The values of $\rho_p$ can be directly related to the potential. Consider first the
inhomogeneous equation for any other non-vanishing field $\phi_{k\gamma}$
\beqx
\chi_{k\gamma}  =  r^2\left[ \lambda_1  \chi_{k\gamma}\left(\chi_{j\beta} \chi^*_{j\beta}\right) +  \lambda_2 \chi_{k\beta}\chi_{j\gamma}\chi^*_{j\beta}
+  \lambda_3 \chi_{k\beta} \chi_{j\tilde\beta}\chi_{j\tilde\gamma}^* 
 +    \lambda_4 \chi_{j\gamma} \chi_{j\beta}\chi_{k\beta}^*+  \lambda_5^* \chi^*_{j\tilde\gamma}   \chi^*_{j\beta} \chi^*_{k\tilde\beta}\right]
\eeqx
By subtracting (\ref{HomogenTwo}) times $\chi_{k\gamma}$ and  using the principle that cancellations depending on special relations among the $\lambda$'s are not acceptable,
we get
\beqa\label{RhoEq}
\sum_{j,\beta}{\chi_{k\beta}\chi_{j\gamma}\chi^*_{j\beta}}&=&\rho_2{\chi_{k\gamma}}\\
\sum_{j,\beta}\chi_{k\beta} \chi_{j\tilde\beta}\chi_{j\tilde\gamma}^*&=&\rho_3{\chi_{k\gamma}}\\
\sum_{j,\beta}\chi_{j\gamma} \chi_{j\beta}\chi_{k\beta}^*&=&\rho_4{\chi_{k\gamma}}\\
\sum_{j,\beta}\chi^*_{j\tilde\gamma}   \chi^*_{j\beta} \chi^*_{k\tilde\beta}&=&\rho_5{\chi_{k\gamma}}
\eeqa
These equations are obtained here for $\chi_{k\gamma} \not =0$, but it also hold for $\chi_{k\gamma} =0$, because then they are just the 
homogeneous equations. 
Now we multiply both sides of these relations with $\chi_{k\gamma}^*$ and sum over $k$ and $\gamma$. Then we get, in terms of the potentials $V_p$ defined in 
(\ref{PotentialTwo})
\beqa
\rho_p&=&\frac{V_p}{\mu^4 r^4 P} \quad \hbox{for}\ p=2,3,4\\
\rho_5&=&\frac{V^*_5}{\mu^4 r^4 P} \
\eeqa
These expressions show in particular that $\rho_p$ is gauge invariant, which was not manifest in the construction we gave above.
Note that $\rho_2$ and $\rho_3$ are manifestly real because of their definition (\ref{RhoDef}). Furthermore, $\rho_4$ is proportional to $V_4$, which is manifestly real, but this
proportionality holds only for solutions of the equation of motion.  The coupling constants $\lambda_2$, $\lambda_3$  and $\lambda_4$ are real, and since $Q$ must be real this implies that
$\rho_5\lambda_5^*$ must be real, although in general neither $\rho_5$ nor $\lambda_5$ are real themselves. This implies
\beqx
\lambda_5^* \rho_5=\pm |\lambda_5|| \rho_5|
\eeqx

Hence for solutions to the equations of motion $\lambda_5 V_5$ is real.
Therefore 
\beqx
\lambda_5^*V_5^*=\tfrac12 (\lambda_5 V_5 +\lambda_5^*V_5^*)
\eeqx
Using this result and(\ref{PotentialTwo})  we can  express the entire quartic contribution to the potential in  terms of the $\rho$-parameters and $P$:
\beqa
V&=&-\mu^2  \phi_{i\alpha} \phi^*_{i\alpha}+  \tfrac12  \mu^4 r^4 P\left( \lambda_1 P
+     \lambda_2  \rho_2
+   \lambda_3 \rho_3   +  \lambda_4  \rho_4
+     \lambda_5^* \rho_5\right)\\
&=&-\mu^4 r^2 P + \tfrac12  \mu^4 r^4 P(\lambda_1 P+Q)\\
&=&-\tfrac12\mu^4 r^2 P 
\eeqa
where in the last step we used (\ref{rEquation}). This result is consistent with the general formula (\ref{VacEn}).

\subsubsection{Disjoint solutions}\label{Disjoint}

If one considers Higgs potentials for $G(N) \times H(M)$ group combinations, solutions must exist already for the smallest allowed
values of $N$ and $M$: if $\mu^2 < 0$, then $\phi=0$ is not a minimum, and hence if the potential is bounded there must exist a non-trivial minimum.
These solutions remain valid if one enlarges $N$ and $M$ and chooses all additional elements of $\phi_{i\alpha}$ to be zero. 

Now consider two such solutions to the equations of motion, $\phi^A$ and $\phi^B$. 
One may attempt to combine two or more solutions, by choosing disjoint block sub-matrices of  $\phi_{i\alpha}$ and embedding a known solution in it.
Here  by ``disjoint" we mean first of all that no rows or columns exist with non-zero elements of both  $\phi^A$ and $\phi^B$. But we need a slightly more general
notion of non-overlapping, namely one that includes the effect of the matrices $D_{ij}$ or $C_{\alpha\beta}$.  In practice these matrices are either diagonal, or block-diagonal
in terms of $2\times 2$ blocks, as happens for symplectic groups. In that case we combine rows or columns  into pairs linked by $C$ or $D$, and we extend the notion
of disjoint  to pairs of rows or columns. 

The homogeneous equations are automatically
satisfied for combinations of solutions in disjoint sub-blocks.  This is because non-zero elements of $\phi_{i\alpha}$
only connect indices belonging to the corresponding solution. 

But this is not true for the inhomogeneous equations because of the $\lambda_1$ term. This
includes a sum over all $\phi_{j\beta} \phi^*_{j\beta}$, which changes if solutions are added. 

Consider  two distinct solutions $\chi$ and $\xi$, each satisfying 
\begin{eqnarray}\label{TwoSols}
 \chi_{i\alpha}  &=&   \lambda_1  \chi_{i\alpha} r_1^2 P_1  +   \lambda_2 r_1^2 \chi_{i\beta}\chi_{j\alpha}\chi^*_{j\beta}+ \ldots \NO\\
 \xi_{p\mu}  &=&     \lambda_1  \xi_{p\mu}r_2^2P_2  +   \lambda_2 r_2^2\xi_{p\nu}\xi_{q\mu}\xi^*_{q\nu}+ \ldots 
\end{eqnarray}
where the indices $(i,\alpha)$ and $(p,\mu)$ are disjoint in the way explained above.
Here we use the dimensionless unit introduced above; the two solutions are
\beqa
\phi^1 _{i\alpha} &=&   r_1 \chi_{i\alpha};   \ \ r_1=\sqrt{\frac{1}{P_1 \lambda_1 + Q_1}}\\
\phi^2 _{i\alpha} &=&   r_2 \xi_{i\alpha};   \ \ r_2=\sqrt{\frac{1}{P_2 \lambda_1 + Q_2}}
\eeqa
with 
\beqx
P_1=\sum_{i,\alpha} \chi_{i\alpha} \chi^*_{i\alpha}\ , \ \ \  \ P_2=\sum_{i,\alpha} \chi_{p\mu} \chi^*_{p\mu}
\eeqx
This combined solution satisfied the homogeneous equations, but the inhomogeneous ones become
\begin{eqnarray}\label{TwoSols}
 \chi_{i\alpha}  &=&   \lambda_1  \chi_{i\alpha} (r_1^2 P_1 +r_2^2P_2) +   \lambda_2 r_1^2 \chi_{i\beta}\chi_{j\alpha}\chi^*_{j\beta}+ \ldots \NO\\
 \xi_{p\mu}  &=&     \lambda_1  \xi_{p\mu}(r_1^2 P_1 +r_2^2P_2)  +   \lambda_2 r_2^2\xi_{p\nu}\xi_{q\mu}\xi^*_{q\nu}+ \ldots 
\end{eqnarray}
Since the homogeneous 
equations are invariant under a simultaneous rescaling of all fields one may hope that we can solve these equations by a rescaling
\beqx
r_1 \ \rightarrow\ S_1 r_1; \ \ \ \ r_2 \ \rightarrow\ S_2 r_2
\eeqx
Now the inhomogeneous equations are
\begin{eqnarray}\label{RescaledEquations}
\chi_{i\alpha} &=&   \lambda_1  \chi_{i\alpha}(S_1^2 r_1^2 P_1+S_2^2 r_2^2P_2) + S_1^2 r_1^2\lambda_2 \chi_{i\beta}\chi_{j\alpha}\chi^*_{j\beta}+\ldots\NO\\
 \xi_{p\mu}  &=&   \lambda_1  \xi_{p\mu}(S_1^2r_1^2 P_1+S_2^2 r_2^2 P_2) + S_2^2r_2^2 \lambda_2 \xi_{p\nu}\xi_{q\mu}\xi^*_{q\nu}+\ldots
\end{eqnarray}
Now we subtract $S_1^2$ times the single solution equation of motion (\ref{TwoSols}) from the first, and analogously for the second. Then the $ \lambda_p$-terms for $p \geq 2$ cancel, and
we
can divide by $\chi_{i\alpha}$ resp.  $\xi_{p\mu}$. The we solve for $S_1^2$ and $S_2^2$ and we get
\beqx
S_1^2 = \frac{ 1-    \lambda_1 r_2^2 P_2}{  1 -   \lambda_1^2 r_1^2r_2^2 P_1P_2}\quad\quad\quad\quad
S_2^2 = \frac{ 1-   \lambda_1r_1^2 P_1}{ 1 -    \lambda_1^2r_1^2r_2^2 P_1P_2}
 \eeqx
Note that if we combine two identical solutions $(r_1=r_2, P_1=P_2)$ we find
\beqx
S_1^2=S_2^2=\frac{1}{1+\lambda_1r^2 P}
\eeqx
so that the $r$ parameter $r_{\rm comb}$ of the combined solution is
\beqx
S_1^2r_2=S_2^2r_2^2=r_{\rm comb}^2=\frac{1}{2P\lambda_1+Q}
\eeqx
It is easy to show that this process can be continued, and that for a combination of $K$ identical solutions the result is
\beqx
r^2=\frac{1}{KP\lambda_1+Q}
\eeqx
This can be verified by working out the combination of two multi-solutions, one built out of $K_1$ and one build out of $K_2$ basic solutions. We can do
that in general, for one $K_1$-fold solution with parameters $P_1$ and $Q_1$, and one $K_2$-fold solution with parameters $P_2$ and $Q_2$.

We find
\begin{equation}\label{CombinedScale}
 S_1^2 r_1^2
=\frac{Q_2} {  \lambda_1( K_2P_2Q_1+K_1P_1Q_2) + Q_1Q_2}
\end{equation}
with an analogous formula for $S_2^2 r_2^2$, with ``1" and ``2" interchanged. We see that if $P_1=P_2$ and $Q_1=Q_2$, then indeed we get the the expected result for a 
$(K_1+K_2)$-fold solution.
Furthermore we see then that $S_1^2r_1^2=S_2^2r_2^2$  if and only if $Q_1=Q_2$. For generic $\lambda_i$ we can
have equality of $Q_1$ and $Q_2$ if and only if $\rho_i^{(1)}=\rho_i^{(2)}$ (the upper index labels the solution).

Note that even for distinct solutions, there is always a solution for the scale factors $S_1$ and $S_2$. But now we can consider the rescaled equations (\ref{RescaledEquations}) for
the special normalizing fields  we have chosen to define $\rho_i$. Theses are the field $\chi_{11}=1$ chosen earlier, and the analogous choice for $\xi$. We can divide the
equations by these fields and subtract them. Then we find
\beqx
S_1^2 r_1^2 \rho_i^{(1)}=S_2^2 r_2^2 \rho_i^{(2)}
\eeqx
Using Eqn (\ref{CombinedScale}) we can write this as
\beq\label{RhoForm}
Q_2 \rho_i^{(1)}=Q_1 \rho_i^{(2)}
\eeq
Note that if we contract this with $\lambda_i$, summing over $i=2\ldots 5$, this is automatically satisfied.
Without summation, these equations imply that two solutions can only be combined if their values of $\rho_p$ for different $p$ have the same ratios. 
\beqx
\frac{\rho_i^{(1)}}{\rho_i^{(2)}}=\frac{Q_1}{Q_2}
\eeqx
Hence in particular
they must be simultaneously zero. 
Furthermore we can divide  (\ref{RhoForm}) on both sides by the $i=2$ equation, which is always non-trivial. This implies that
for two solutions to be combined, one must have
\beqx
\rho_p^{(2)}=\left[\frac{\rho_2^{(1)}}{\rho_2^{(2)}}\right]{\rho_p^{(1)}}
\eeqx
Although this still allows a common scaling, as we will see there are no two cases where the $\rho_p$ parameters differ only by a common scale.

\subsubsection{The main argument}

Now we combine all the foregoing results. As already discussed in section \ref{InhomEq},
we can always rotate the first row to the form 
\beqx
(\phi_{11},0,\ldots,0)
\eeqx
Then the homogeneous equation for $G\times H$ implies that all columns are orthogonal to column 1. Now row 1 is fixed, but we have $U(N-1)$ or $O(N-1)$ rotations at our disposal
to clean up the last $N-1$ entries of column 1. 

Consider first  unitary column transformations.  Then we can set $\phi_{j1}=0$ for $j \geq 3$, and make $\phi_{21}$ real.  
Since all remaining columns must be orthogonal to column 1, this
implies that $\phi_{2\alpha}=0$ for $\alpha \geq 2$. Now the first two rows of the matrix have the form
\beqx
\begin{pmatrix}
\phi_{11}&0&\ldots&0\\
\phi_{21}&0&\ldots&0
\end{pmatrix}
\eeqx
If $\phi_{21}=0$, we have obtained a matrix where $\phi_{11}$ is the only non-zero element in the first row and column. If 
$\phi_{21}\not=0$, the fact that the remainder of the first two rows vanishes  implies that  we can apply a $U(2)$ rotation to the first column, and rotate $\phi_{21}$ to zero.  Hence one again we end up with an element $\phi_{11}$ in an otherwise vanishing row and column.  
In either case the result is  a disjoint $1\times 1$ block matrix if $H(M)=U(M)$. Note that even though the matrix is the same if  $H(M)=USp(M)$, it is not necessarily disjoint, since this would require column 2 to vanish.

In the $O(N)$ case the argument is similar. 
Now we rotate the last $N-1$ rows so that the first column has one of the following three features
\beqa
{\bf{a:}}\quad\quad\phi_{j1}&=&0\quad \hbox{for}\ j \geq 2\\
{ \bf{b:}}\quad\quad \phi_{j1}&=&0\quad \hbox{for}\ j \geq 3; \quad \phi_{21}\not=0;\\
{\bf{c:}}\quad \quad \phi_{j1}&=&0\quad \hbox{for}\ j \geq 4; \quad \phi_{21}\not=0;\quad \phi_{31}\not=0;\quad \phi_{21}\phi_{31}^*-\phi^*_{21}\phi_{31}\not=0
\eeqa
Note that the last condition is that $\phi_{21}$ and $\phi_{31}$ have a different phase. If that were not the case we could rotate $\phi_{31}$ into $\phi_{21}$ and then case {\bf c} turns into case {\bf b}. 
In case {\bf a} we have a disjoint $1\times 1$ block matrix if $H(M)=U(M)$ and the discussion is as before. 
In case {\bf b}
the general orthogonality equation (\ref{OrthoOne}) implies that $\phi_{2\alpha}=0$  for $\alpha \geq 2$. Then either $\phi_{21}$ is real, and it can be rotated into $\phi_{11}$, or it is not real 
and then the arguments at the end of section \ref{SpecialElt} show that $\phi_{21}=i\phi_{11}$.
In case {\bf c} we use both  (\ref{OrthoOne}) and (\ref{OrthoThree}) plus the fact that  $\phi_{21}$ and $\phi_{31}$ have a different phase to show that 
 that  $\phi_{2\alpha}=\phi_{3\alpha}=0$  for $\alpha \geq 2$. Now we have an $O(3)$ gauge symmetry in the first three rows at our disposal to reduce case {\bf c} to case {\bf b}.
We may now normalize $\phi_{11}$ to 1 by defining the parameter $r$ appropriately. We find then that we have two possible disjoint solutions. One works for both $U(N)$ and $O(N)$, and
is characterized by an upper left $1\times 1$ block 
\beqx
{\rm A}:\quad \begin{pmatrix}
1
\end{pmatrix}
\eeqx
The other holds only for $O(N)$ and is characterized by an upper left $2\times 1$ block 
\beqx
{\rm B}:\quad  \begin{pmatrix}
1\\
i
\end{pmatrix}
\eeqx

If the column group $H(M)$ is $U(M)$ these blocks are really disjoint, and we can repeat the process for the block matrix defined by the last $M-1$ column and the last $N-1$ or $N-2$ rows.
This sub-matrix is treated in exactly the same way, and yields the same solutions. From the general argument in section \ref{Disjoint} we know that these solutions can only be combined with the upper left
block if they are identical, or zero. This follows from the fact that their parameters $\rho_p$ are not proportional. The parameters are shown in table \ref{RhoTable}. We repeat this process until the remaining lower 
right block matrix is identically zero.

If the column group is $USp(M)$ we start in the same way. However, now the upper left blocks are not strictly disjoint from the rest of the matrix, because $USp(N)$ links columns 1 and 2.
We can deal  with the cases   $G(N)=U(N)$ and $G(N)=O(N)$ simultaneously. The first step yields a $r\times 1$ upper left block, where $r=1$ or $2$.
Now we clean up column 2 using $G(N-r)$ rotations in the  last $N-r$ rows. Note that in those last $N-r$ rows all elements to the left of column 2 are zero, and all elements to the
right of column 2 are arbitrary complex numbers, so we can freely use $G(N-r)$ transformations acting on column 2.

We get essentially the same three options as above. Option {\bf a} is that column 2 vanishes completely. Then the upper block is
disjoint in the  $USp(N)$ sense. Option {\bf b} is that column 2 contains just one non-zero element, $\phi_{r+1,2}$. Then the  homogeneous equations (\ref{OrthoTwo}) tells us that the remainder of row 2 to
the right of $\phi_{r+1,2}$ must vanish, and this makes the entire block disjoint in the $USp(N)$ sense. Option {\bf c} is that column 2 contains two non-vanishing elements. This can only happen
if $G(N)=O(N)$ and if those two elements have different phases. Now we use both (\ref{OrthoTwo}) and (\ref{OrthoFour}) (which indeed is available for $O(N)$) to show that rows $r+1$ and $r+2$ are
zero except on column 2. We conclude that in all cases the $r\times s$ upper left blocks are completely disjoint from the rest of the matrix. So now we can continue the argument in the last $N-r$ rows, and
$N-s$ columns.

We have now reached a situation where both column 1 and 2 consist entirely of pivot elements, and we can apply the results of section \ref{SpecialElt}. This tells us
that $\phi_{42}$ must be equal to $i\phi_{32}$ and that  columns with one and two pivot elements cannot be combined. This leaves us with the following four possibilities
for the upper left block
\beqx
{\rm A}_0: \begin{pmatrix}
1& 0
\end{pmatrix}  \quad
{\rm C}_x: \begin{pmatrix}
1& 0\\
0 & x
\end{pmatrix}
\eeqx
\beqx
{\rm B}_0: \begin{pmatrix}
1& 0\\
i & 0
\end{pmatrix}  \quad
{\rm D:} \begin{pmatrix}
1& 0\\
i& 0\\
0 & x\\
0 & ix
\end{pmatrix}
\eeqx
Here $x$ is a phase, since we know from (\ref{InhomOne}) that all columns must have equal norm.  All four must be considered for $O(N)\times USp(N)$ and only $A$ and $C$ for $U(N)\times USp(N)$. 

The value of $x$ requires some additional discussion. Consider first ${\rm C}_x$. If $G(N)=U(N)$ we can make $x$ real using a phase rotation, and then the equations of motion guarantee that $x=1$. But this is
not true if $G(N)=O(N)$. In that case there is a $\lambda_5$ term in the potential, and the quantity $\rho_5$ exists, and is equal to $(x^*)^2$ (see \ref{RhoDef}). Then the condition that $\lambda_5\rho_5^*$ be real 
determines $x$ up to a factor $i$. The solutions are
\beq\label{xlambda}
x=\sqrt{\frac{\lambda^*_5}{|\lambda_5|}} \equiv y;\quad \quad\quad \quad x=i\sqrt{\frac{\lambda^*_5}{|\lambda_5|}}=iy
\eeq
In each case there are two roots, but using $O(N)$ we can change the sign of row 2, and map them to each other.

In case  D  we can combine an orthogonal $SO(2)$ rotation  on the first two rows with a diagonal phase rotation on column 1 and 2 in $SU(2)$ to obtain
\beqx
\begin{pmatrix}
(c+is)e^{i\theta} & 0\\
(-s+ic)e^{i\theta} & 0\\
0 & xe^{-i\theta}\\
0 & ixe^{-i\theta}
\end{pmatrix}
\eeqx
Now we can choose $\theta$ to cancel the phase of $x$, and choose $c$ and $s$ to cancel $\theta$ in $\chi_{11}$, so that the final result is 
\beqx
D: \begin{pmatrix}
1& 0\\
i& 0\\
0 & 1\\
0 & i
\end{pmatrix}
\eeqx
In table \ref{RhoTable} we summarize all solutions. The table is organized in terms of four vertical blocks, that respectively specify the original intersecting brane group,
the values of the parameter $P$ and $\rho_p$, the basic block matrix $\mathbbm{X}$ and its size, and the unbroken subgroup and its embedding. The latter will be
discussed in the next section.
Note that, as announced earlier, for a given combination of $G$ and $H$ there are no two cases 
with vectors $\rho_p$ that are proportional to each other. Therefore a general solution is a combination of $K$ identical basic blocks, and never a combination of
different blocks.

 \begin{table}[h]
\begin{center}
\begin{tabular}{|c||c|c|c|c|c||c|c|c||c|c|} \hline
 Group &             $P$  & $\rho_2$  & $\rho_3$ & $\rho_4$ & $\rho_5$ & $\mathbbm{X}$ &$p$ &$q$ & Subgroup & Emb.\\ \hline
 $U(N)\times U(M)$ &  $1$  & $1$          & $-$         & $-$         & $-$ & ${\rm A}$  &$1$ &$1$& $U(K)$ & (0,0)\\  
$O(N)\times O(M)$ &  $1$  & $1$          & $-$         & $-$         & $-$ & ${\rm A}$  &$1$ &$1$& $O(K)$ &(0,0) \\  
$USp(N)\times USp(M)$ &  $2$  & $2$          & $-$         & $-$         & $-$ & ${\rm C}_1$  &$2$ &$2$&$USp(2K)$& (0,0)\\  
  $O(N)\times U(M)$ &  $1$  & $1$          & $-$ & $1$ & $-$ & ${\rm A}$  &$1$ &$1$&$O(K)$ & (0,1)\\  
  $O(N)\times U(M)$ &   $2$ & $2$          & $-$ & $0$ & $-$ &${\rm B}$ &$2$ &$1$  &$U(K)$ & (3,0)\\  
   $U(N)\times USp(M)$ &  $1$ & $1$         & $0$ & $-$ & $-$ & ${\rm A}_0$  &$1$ &$2$& $U(K)$ & (0,4)\\  
  $U(N)\times USp(M)$ &       $2$ & $1$         & $1$ & $-$ & $-$ & ${\rm C}_1$ &$2$ &$2$  & $USp(2K)$ &(2,0)\\  
   $O(N)\times USp(M)$ &  $1$ & $1$            &  $0$ & $1$ & $0$ & ${\rm A}_0$  &$1$ &$2$& $O(K)$ & (0,5)\\  
  $O(N)\times USp(M)$ & $2$ & $2$              & $0$ & $0$ & $0$ &${\rm B}_0$  &$2$ &$2$ & $U(K)$   &(3,4)\\  
 $O(N)\times USp(M)$ &  $2$ & $1$             & $1$ & $1$ & $\omega_5$ & ${\rm C}_y$ &$2$ &$2$ & $U(K)$ &(3,4) \\  
  $O(N)\times USp(M)$ &  $2$ & $1$             & $1$ & $1$ & $-\omega_5$ & ${\rm C}_{iy}$ &$2$ &$2$ & $U(K)$ &(3,4)\\  
   $O(N)\times USp(M)$ &  $4$ & $2$             & $2$ & $0$ & $0$ & D &$4$ &$2$ & $USp(2K)$ &(6,0) \\  
  \hline
\end{tabular}
\caption{All solutions and the resulting subgroup embedding. In column 6, $\omega_5 = \lambda_5/|\lambda_5|$.}\label{RhoTable}
\end{center}
\end{table}

Note that the subscript of the matrix {\rm C} denotes the value of $x$, but the subscripts on A and B have a different purpose: they indicate that the second column 
vanishes. This would be irrelevant if $H(m)=U(m)$,  but it is needed in $USp(M)$ in order for the block to be disjoint from the rest of the matrix.

\subsection{Subgroups}

Now we will determine the subgroups that are left unbroken by these solutions.
 Since only identical blocks can be repeated, the general form of the vacuum is
\beqx
v \mathbbm{1}_K\otimes \mathbbm{X}
\eeqx
where $\mathbbm{X}$ denotes the blocks in the last column of the table and $\mathbbm{1}_K$ is
the $K\times K$ unit matrix.

We begin with a list of all subgroup embeddings that occur. First of all we need the ``brane separation" embeddings
\begin{eqnarray}\label{BraneSplit}
U(k+\ell) &\rightarrow& U(k)\times U(\ell)\NO\\
O(k+\ell) &\rightarrow& O(k)\times O(\ell)\\
USp(2k+2\ell) &\rightarrow& USp(2k)\times USp(2\ell)\NO
\end{eqnarray}
In all these cases the vector representation splits as $(V,1)+(1,V)$. This embedding is used to split off the group  $G(N-Kp)\times H(M-Kq)$ that acts trivially on the vacuum. This part of the breaking requires no further discussion, so we leave out the  $G(N-Kp)\times H(M-Kq)$ factor of the unbroken subgroup henceforth.

 Further discussion will be needed to determine which parts of $G(Kp)\times H(Kq)$ survive, but roughly speaking it will
be some diagonal subgroup of the two factors, obtained by means of a suitable left-right combination of one of the embeddings listed
in table \ref{SubgroupTable}. 
\begin{table}[h]
\begin{center}
\begin{tabular}{|c|c|c|c|} \hline
nr. & Group &      Subgroup    & Vector decomposition  \\ \hline
0& $G(p)$& $G(p)$    & V     \\  
1& $U(p)$& $O(p)$    & V     \\  
2&$U(2p)$& $USp(2p)$    & V     \\  
3&$O(2p)$& $U(p)$    & V+V$^*$     \\  
4&$USp(2p)$& $U(p)$   & V+V$^*$     \\  
5&$USp(2p)$& $O(p)$    & 2V     \\  
6&$O(4p)$& $USp(2p)$    & 2V     \\  
     \hline
\end{tabular}
\caption{Basic subgroup embeddings.}\label{SubgroupTable}
\end{center}
\end{table}
By embedding 0 we mean the trivial one, available for $U$, $O$ and $USp$. 
Embedding 1 is simply the restriction from complex matrices to real ones. Embedding 2 is the restriction from complex to quaternionic, {\it i.e.} unitary 
matrices are restricted to the subset $U^*=hU h^T$  where $h$ is anti-symmetric.  
Embedding 3 is a well-known one, used in GUT theories
for embedding $SU(5)$ GUTs in $SO(10)$; for further details see the appendix. Embedding 4 is similar, and follows immediately from the standard basis used for
symplectic groups, as explained in the appendix. Embedding 5 is obtained by combining 4 and 1, and embedding 6 by combining 3 and 4. 

Embedding 3 is best understood by extending $O(2K)$ to $U(2K)$ and then conjugate
the entire $O(2K)$ group within $U(2K)$. The resulting group matrices  are not  real, but this is as good a definition of $O(2K)$ 
as the standard one. Moreover, we can always transform the result back to the real form, if we wish. For this
transformation we use the matrix (\ref{XTransform}), but with rows and columns rearranged into pairs, exactly 
as in the symplectic case (as explained in appendix \ref{QuatBas}). The matrix $Z$ now takes the form
\beq\label{XXtransform}
Z=\frac{1}{\sqrt2}
\mathbbm{1}_K \otimes 
\begin{pmatrix}
1 & 1 \\
i  & -i
\end{pmatrix}
\eeq
We transform the orthogonal group generators $O$ to $\tilde O=Z^{\dagger}O Z$, and multiply the vacuum matrix on the right 
by $Z^{\dagger}$. The advantage of this basis becomes clear when we make $O(N)$ act on vacua of the form {\bf B} and {\bf D}, which 
have column vectors $(1,i)$. In the new basis these take the form $(1,0)$.
There is a subgroup $U(K) \subset O(2K)$ that acts on the odd indices as a unitary matrix $Y$, and on the
even indices as  $Y^*$. There are additional generators in $O(2K)$, but they map the odd components to the even ones, and
this can never be repaired by a transformation acting on the columns. 

The complexified $O(N)$ basis reveals some nice analogies between the symplectic and orthogonal cases, but  it is probably not preferable to work
in that basis from the start. First of all this basis is only useful for even $N$, and secondly the concept of disjoint matrices becomes less convenient 
in the complexified basis. Basis elements now come in pairs, as for $USp(N)$, because the metric is build out of $2\times2$ $\sigma_1$ blocks. Hence
only pairs can be disjoint. Another way of saying this is that the complexified basis is good for solutions of type {\bf B} and {\bf D}, but inconvenient for
type {\bf A}  (as well as {\bf C}).

We will specify for each case the embedding of the subgroup in $G(Kp)\times H(Kq)$. This will be done by specifying a pair of labels $(m,n)$ that each
refer to a line in table \ref{SubgroupTable}. From $(m,n)$ one can determine the embedding in the vector representations of the two groups $G(Kp)$ and $H(Kq)$. 
From this we derive the decomposition of the Higgs field itself, which must include a singlet, corresponding
to the vacuum expectation value.

\paragraph{$\bullet$ $U(N)\times U(M)$, Type A.}

The vacuum, limited to the $K\times K$ subspace where the v.e.v. is non-trivial, is a multiple of the $K\times K$ unit matrix.
This case was already discussed in  \cite{Li:1973mq}. We just use it to illustrate our notation.  
\beqa
U(K)\times U(K) &\rightarrow& U(K) \quad\quad \hbox{embedding}\ (0,0)\\
({\rm V},{\rm V}) &\rightarrow&  1+{\rm Adj}_K
\eeqa
Here ${\rm Adj}_K$ is the irreducible $U(K)$ representation of dimension $K^2-1$, the  adjoint representation 
of $SU(K)$ subgroup of $U(K)$, with $U(1)$-charge 0.
The first component of the decomposition of the Higgs field, (V,V), is the singlet that corresponds to the Higgs v.e.v.  

\paragraph{$\bullet$  $O(N)\times O(M)$ Type A.}

This was also discussed in \cite{Li:1973mq}. The result is 
\beqa
O(K)\times O(K) &\rightarrow& O(K) \quad\quad \hbox{embedding}\ (0,0) \\
({\rm V},{\rm V}) &\rightarrow&  1+{\rm A}+{\rm S}
\eeqa
In this case the Higgs singlet comes out of the trace of the symmetric tensor.

\paragraph{$\bullet$ $USp(2N)\times USp(2M)$ Type C$_1$.}

In terms of quaternions, the vacuum has the same form as the previous two examples. It is proportional to a $K\times K$ diagonal 
matrix of unit  quaternions. The reason it appears here as type C$_1$ rather than A is that we have written the quaternions in a complex base of twice the size. In terms
of complex fields the vacuum is 
\beqx
v
\mathbbm{1}_K \otimes 
\begin{pmatrix}
1 & 0 \\
0 & 1
\end{pmatrix}
\eeqx
Note that a diagonal with an odd number of non-vanishing entries does not even respect the quaternionic condition, so it cannot occur. 
The result is
\beqa
USp(2K)\times USp(2K) &\rightarrow& USp(2K)\quad\quad \hbox{embedding}\ (0,0) \\
({\rm V},{\rm V}) &\rightarrow&  1+{\rm A}+{\rm S} 
\eeqa
The only differences with orthogonal case is that all dimensions are even, and that the anti-symmetric representation is reducible: a symplectic 
trace must be removed. In the orthogonal case the symmetric representation is the one that must be made traceless. In both cases, the trace
provides the Higgs representation.

\paragraph{$\bullet$ $O(N)\times U(M)$ Type A.}

In this case the vacuum block matrix $\mathbbm{X}$ is $(1)$. This is very similar to the $U(N)\times U(M)$ and $U(N)\times U(M)$ breakings, Within
the right $U(K)$ subgroup only the $O(K)$ transformations, the real subgroup of $O(K)$  can be compensated by orthogonal transformations. 
The remaining $U(K)$ transformations, acting infinitesimally, generate imaginary parts that cannot be removed by an orthogonal transformation.
Therefore we get
\begin{eqnarray*}
O(K)\times U(K) &\rightarrow& O(K)\quad\quad \hbox{embedding}\ (0,1)\\
({\rm V},{\rm V}) &\rightarrow&  1+{\rm S}+{\rm A}
\end{eqnarray*}

\paragraph{$\bullet$ $O(N)\times U(M)$ Type B.}

If we use $K$ basic blocks B, then we get a vacuum matrix that can be brought into the form
\beqx
v
\mathbbm{1}_K \otimes 
\begin{pmatrix}
1  \\
i 
\end{pmatrix}
\eeqx
The part of $O(N)$ that is affected by the v.e.v. is $O(2K)$. The determination of the symmetry of the vacuum can be done most efficiently
by using the transformation (\ref{XXtransform}). Clearly the left $U(K) \subset O(2K)$ can be undone by
a right $U(K)$ transformation. Hence the final result is

\beqa
O(2K)\times U(K) &\rightarrow& U(K)\quad\quad \hbox{embedding}\ (3,0)\\
({\rm V},{\rm V}) &\rightarrow&  1+{\rm Adj}_K+{\rm A}+{\rm S} 
\eeqa

\paragraph{$\bullet$ $U(N)\times USp(M)$, Type ${\bf A}_0$.}

The block matrix $\mathbbm{X}$ that defines the vacuum is
\beqx
\mathbbm{X}=
\begin{pmatrix}
1 & 0 \\
\end{pmatrix}
\eeqx
With $K$ diagonal copies of that matrix, the effect is to break the $SU(2)$ factors acting on each column pair, so that only $U(K)$ remains. This combines with 
a $U(K)$ factor that acts on the row index. The result is
\beqa
U(K)\times USp(2K) &\rightarrow& U(K)\quad\quad \hbox{embedding}\ (0,4)\\
({\rm V},{\rm V}) &\rightarrow&  1+{\rm Adj}_K+{\rm A}+{\rm S}
\eeqa
Note that this is like the mirror image  of case B for $O(N)\times U(M)$ discussed above, after using the $Z$-transformation (\ref{XXtransform}).

\paragraph{$\bullet$ $U(N)\times USp(M)$, Type C$_1$.}

The block matrix $\mathbbm{X}$ is the $2\times 2$ unit matrix. Clearly, if we act on $K$  of these blocks with $USp(2K)$ from the left and the right,
the diagonal combination is preserved. Hence we get
\beqa
U(2K)\times USp(2K) &\rightarrow& USp(2K)\quad\quad \hbox{embedding}\ (2,0)\\
({\rm V},{\rm V}) &\rightarrow&  1+{\rm A}+{\rm S}
\eeqa

\paragraph{$O(N)\times USp(M)$, Type ${\bf A}_0$.}
The vacuum is
\beqx
\mathbbm{X}=
\begin{pmatrix}
1 & 0 \\
\end{pmatrix}
\eeqx
The affected part of the group is $O(K)\times USp(2K)$. 
 The $O(K)$ subgroup of $O(N)$ can keep the vacuum invariant if it is combined with
a $O(K)$ rotation of the $K$ $1\times 2$ blocks. Hence we need to break $USp(2K)$ acting on those blocks first to 
$U(K)$ and then to $O(K)$. Hence we get 
\beqa
O(K)\times USp(2K) &\rightarrow& O(K)\quad\quad \hbox{embedding}\ (0,5)\\
({\rm V},{\rm V}) &\rightarrow& 2 \times (1+{\rm A}+{\rm S})
\eeqa
This subgroup $O(K)  \subset USp(2K)$  is the maximal subgroup that can survive. The $SU(2)^K$ subgroups of $USp(2K)$ all change
the vector $(1,0)$, and this can never be undone by a $O(K)$ transformation on the row indices. Likewise, the complex transformations in $U(K) \subset USp(2K)$
make the vacuum complex, and then an orthogonal transformation cannot make them real again.

\paragraph{$\bullet$ $O(N)\times USp(M)$, Type ${\bf B}_0$.}

In this case the vacuum has the form
\beqx
v \mathbbm{1} \otimes 
\begin{pmatrix}
1 & 0\\
i  & 0
\end{pmatrix}
\eeqx
The discussion is very similar to case ${\bf B}$ of $O(N) \times U(M)$ combined with case ${\bf A}$ of $U(N) \times USp(M)$. 
The surviving symmetry group is the diagonal combination of the $U(K)$ subgroup of $O(2K)$ and the 
analogous $U(K)$ subgroup of $USp(2K)$.
\beqa
O(2K)\times USp(2K) &\rightarrow& U(K)\quad\quad \hbox{embedding}\ (3,4)\\
(V,V) &\rightarrow& 2 \times (1+ {\rm Adj}_K) + {\rm A}+ {\rm S}+   {\rm A}^*+  {\rm S}^*
\eeqa

\paragraph{$\bullet$ $O(N)\times USp(M)$, Type C.}

The matrix 
\beqx
\begin{pmatrix}
1 & 0\\
0 & x
\end{pmatrix}
\eeqx
is in general complex, and cannot be made real by gauge transformations. Writing $x=e^{i\xi}$ we have
\beqx
\begin{pmatrix}
1 & 0\\
0 & e^{i\xi}
\end{pmatrix}
\simeq
\begin{pmatrix}
e^{i\xi/2} & 0\\
0 & e^{i\xi/2}
\end{pmatrix}
=
e^{i\xi/2}
\begin{pmatrix}
1 & 0\\
0 & 1
\end{pmatrix}
\eeqx
where $\simeq$ denotes gauge-equivalence. The transformation used here is a diagonal $SU(2)$ transformation in $USp(M)$. The overall phase $e^{i\xi/2}$ cannot be transformed away. This is
the only case among the six combinations of groups U, O and S where such a phase can occur. Bi-fundamentals of type  (O,O)  and  (S,S) are real or quaternionic, and  such a phase 
violates these constraints; in any combination that involves a unitary group the phase can be gauged away. So only for (O,S) the phase exists and is not a pure gauge variable. This is also the
reason of existence for the $V_5$ terms in the potential. Without them, overall phase changes of the field $\phi_{i\alpha}$ would give rise to flat directions in the potential. 

We may explore the potential along this phase direction. We get, keeping everything fixed except the phase
\beqx
V(\xi) = \hbox{const}\ + \tfrac12 \lambda_5 R e^{2i\xi} + \lambda_5^* R e^{-2i\xi}=|\lambda_5| R\ \rm{cos}(\eta_5+2\xi)
\eeqx
where $R$ is some real number and $\lambda_5=|\lambda_5| e^{i\eta_5}$. 
As a function of $0 \leq \xi < 2\pi$ the cosine has four extrema, at $\xi=-\tfrac12 \eta_5 + \frac12 \ell \pi, \ell =0,1,2,3$. This implies
\beqx
x=\pm\sqrt{\frac{\lambda^*_5}{|\lambda_5|}};\quad\quad x=\pm i\sqrt{\frac{\lambda^*_5}{|\lambda_5|}}
\eeqx
as we have seen earlier. The signs can be gauged away, so that we end up with two distinct extrema, corresponding to the two options ${\rm C}_x$ and ${\rm C}_{ix}$ in 
table \ref{RhoTable}. Clearly, one of these extrema has more vacuum energy than the other and is therefore a saddle point. The lower of the two can be a saddle point of the full potential, or a local or global minimum, depending
on the values of the other parameters $\lambda_i$.  

In the minimal case, $N=M=2$, the matrix ${\rm C}_1$ breaks the group $O(2)\times USp(2)$ to a diagonal $SO(2)\sim U(1)$, with the $SO(2)$ within $USp(2)\equiv SU(2)$ generated by $\sigma_2$ (the group is
$SO(2)$ rather than $O(2)$ because $SU(2)$ does not contain $O(2)$).
If we replace ${\rm C}_1$ by ${\rm C}_x$, the $SO(2)$ generator is rotated within $SU(2)$ to ${\rm cos}(\xi) \sigma_2 + {\rm sin}(\xi) \sigma_1$.  Hence the $SO(2) \subset SU(2)$ embedding rotates inside
$SU(2)$ as a function of the phase of $\lambda_5$. 
The final results for this embedding are, for both values of the complex parameter $x$
\beqa
O(2K)\times USp(2K) &\rightarrow& U(K)\quad\quad \hbox{embedding}\ (3,4)\\
(V,V) &\rightarrow& 2 \times (1+ {\rm Adj}_K) + A+ S+  A^*+ S^*
\eeqa
Note that this is group-theoretically the same embedding as in case ${\bf B}_0$. However, the subgroup is embedded in a different way in $USp(N)$. This can most easily
be clarified for  the minimal case $N=M=2$. If we work with the real basis for $O(2)$ instead of the complexified basis used in the discussion of case ${\bf B}_0$ then on the $O(2)$ side the action is
identical in both cases. There is only one generator, so we have no choice of embedding. However, in case ${\rm C}_1$  the $O(2)$ action is undone by an $O(2) \subset
USp(2) \sim SU(2)$ that is simply a real restriction of $SU(2)$, whereas in case ${\bf B}_0$ it is undone by an $SU(2)$  group element $e^{i\theta\sigma_3}$.

Although isomorphic subgroups are obtained, these are distinct vacua, with different vacuum energies.

\paragraph{$\bullet$ $O(N)\times USp(M)$, Type D.}

Now the vacuum is
 \beqx
v \mathbbm{1}_K \otimes 
\begin{pmatrix}
1 & 0\\
i  & 0 \\
0 & 1\\
0 & i 
\end{pmatrix}
\eeqx
We use the matrix (\ref{XXtransform}) on the left, to bring the vacuum in the form 
 \beqx
\frac{v}{\sqrt{2}} \mathbbm{1}_K \otimes 
\begin{pmatrix}
1 & 0\\
0  & 0 \\
0 & 1\\
0 & 0 
\end{pmatrix}
\eeqx
On the left the canonical $U(2K)$ subgroup of $O(4K)$ is the only subset that has a chance to be compensate by a transformation
from the right. But we do not have a full $U(2K)$ available on the right; the maximal set of transformations is $USp(2K)$. Hence the
left group must be broken one additional step further to $USp(2K)$. 
  The final result is
\beqa
O(4K)\times USp(2K) &\rightarrow& USp(2K)\quad\quad \hbox{embedding}\ (6,0)\\
(V,V) &\rightarrow& 2\times (1+A+S)
\eeqa
Note the similarity with case ${\bf A}_0$. Indeed, in all results there is a manifest similarity under exchange of orthogonal and symplectic transformations. 
This is also apparent in Table \ref{SubgroupTable}.

\subsection{Comparison of vacuum energies}\label{ComVac}

Now we compare the vacuum energies of the solutions to determine the absolute minimum. This will also provide insight about the reason all these solutions exist.
The vacuum  energy of a multiple solution built out of $K$ disjoint blocks with parameters $P$ and $Q$ is  $-\tfrac12 KP \mu^4 r^2$. This follows from Eqn (\ref{VacEn}) and the
fact that all non-vanishing field values in a solution have the same absolute value. Their total number is the number of blocks, $K$, times the number of non-zero entries per block, $P$. 
This yields
\begin{equation}\label{GeneralEnergy}
E(K,P,Q)=-\frac{KP\mu^4}{2 (KP \lambda_1+Q)}=-\frac{\mu^4}{2 (\lambda_1+Q/KP)}
\end{equation}
The numerator must be positive for any allowed value of $Q, K$ and $P$. We will see in the next section why this must be true, but it is already clear that a solution with negative numerator has
positive vacuum energy, and hence can never be the absolute minimum. If the numerator is positive, the minimal energy is obtained for the minimal value of $Q/KP$. 

For fixed $Q$ and $P$ this implies the following:
If $Q < 0$ the vacuum energy increases with $K$ so that the minimal energy is reached for the smallest non-trivial value of $K$, $K=1$. If $Q>0$ the vacuum energy decreases with $K$, and 
hence the minimum occurs for the largest value of $K$ that is allowed by $M$ and $N$, compared to the sizes $p$ and $q$ of the basic block given in table \ref{RhoTable}. To be precise
\beqx
K_{\rm max}={\rm min}(\lfloor M/p\rfloor ,\lfloor N/q\rfloor ) \ ,
\eeqx
where $\lfloor x \rfloor$ (the ``floor function") is the largest integer smaller or equal to $x$.
Only the values $K=1$ and $K=K_{\rm max}$ can occur as absolute minima
for suitable parameter values. This is in agreement with the results of \cite{Li:1973mq}; in that case $Q=\lambda_2$.

Now we still have to compare different solutions. The ordering of solutions depends on a complicated way on the coupling constants, and it is not worthwhile to
work this out in detail. But it is not difficult to see that -- with one exception, see below -- one can always make choices of $\lambda_2\ldots\lambda_5$ so that $Q$ is negative for one solution, and positive for all others. Then one
can always make $\lambda_1$ large enough so that $\lambda_1+Q/KP > 0$. The solution with negative $Q$ is then the global minimum. This implies that any solution in the table can be an absolute minimum for $K=1$. 

The exception is one of the two solutions with block matrices ${\rm C}_y$ and ${\rm C}_{iy}$. Their values of $Q$, defined in (\ref{Qdef}) are respectively $Q_y=\lambda_2+\lambda_3+\lambda_4+|\lambda_5|$ and 
$Q_{iy}=\lambda_2+\lambda_3+\lambda_4-|\lambda_5|$. Obviously $Q_{iy} < Q_{y}$, and hence only $Q_{iy}$ can be a global minimum, as we have seen already in a different way in the previous section.

The discussion for $K=K_{\rm max}$ is similar. For any solution -- except the two just mentioned -- there is a choice of coupling constants so that its value of $Q/PK_{\rm max}$ value is positive, but smaller than all others. Hence  
any of the solutions in the table, except  ${\rm C}_y$ can occur as a global minimum for $K=K_{\rm max}$.

\subsection{Boundedness and existence of solutions}

Now we discuss two issues that are related: the fact that for certain parameters the potential becomes unbounded from below, and the existence
of singularities in the set of solutions as a function of the couplings.  The parameter $r$, defined  by Eqn. (\ref{rEquation}) must be real, and hence the
argument of the square root must be non-negative. This implies that the quantity $\lambda_1+Q/KP$ must be non-negative. If the numerator is negative for just one solution, one
might conclude that this merely implies that the corresponding solution does not exist, but we will see that in that case the potential is unbounded, so that the other solutions
lose their physical relevance as well.

Consider first what happens if we vary $\lambda_1$, while keeping all other coupling constants fixed. For sufficiently large $\lambda_1$ the potential is bounded, and the 
quantities  $\lambda_1+Q/KP$ are positive for all solutions. If we decrease $\lambda_1$ we reach a singularity at
\beqx
\lambda_1=-\frac{Q}{KP}
\eeqx
The first such singularity we encounter is the one with smallest value of $\frac{Q}{KP}$, which corresponds to the solution that is the global minimum. If we pass through the singularity, the vacuum energy
jumps from $-\infty$ to $+\infty$, and $r$ becomes imaginary. Just before reaching the singularity the energy of the global minimum approaches $-\infty$, indicating that the potential has become unbounded.
If we decrease $\lambda_1$ even more the potential remains unbounded, so that the set $\lambda_1+Q/KP=0$, a hyperplane in the space of all couplings, marks the separation between bounded and 
unbounded potentials.

\begin{figure}[h!]
\begin{center}
\includegraphics[width=9.5cm]{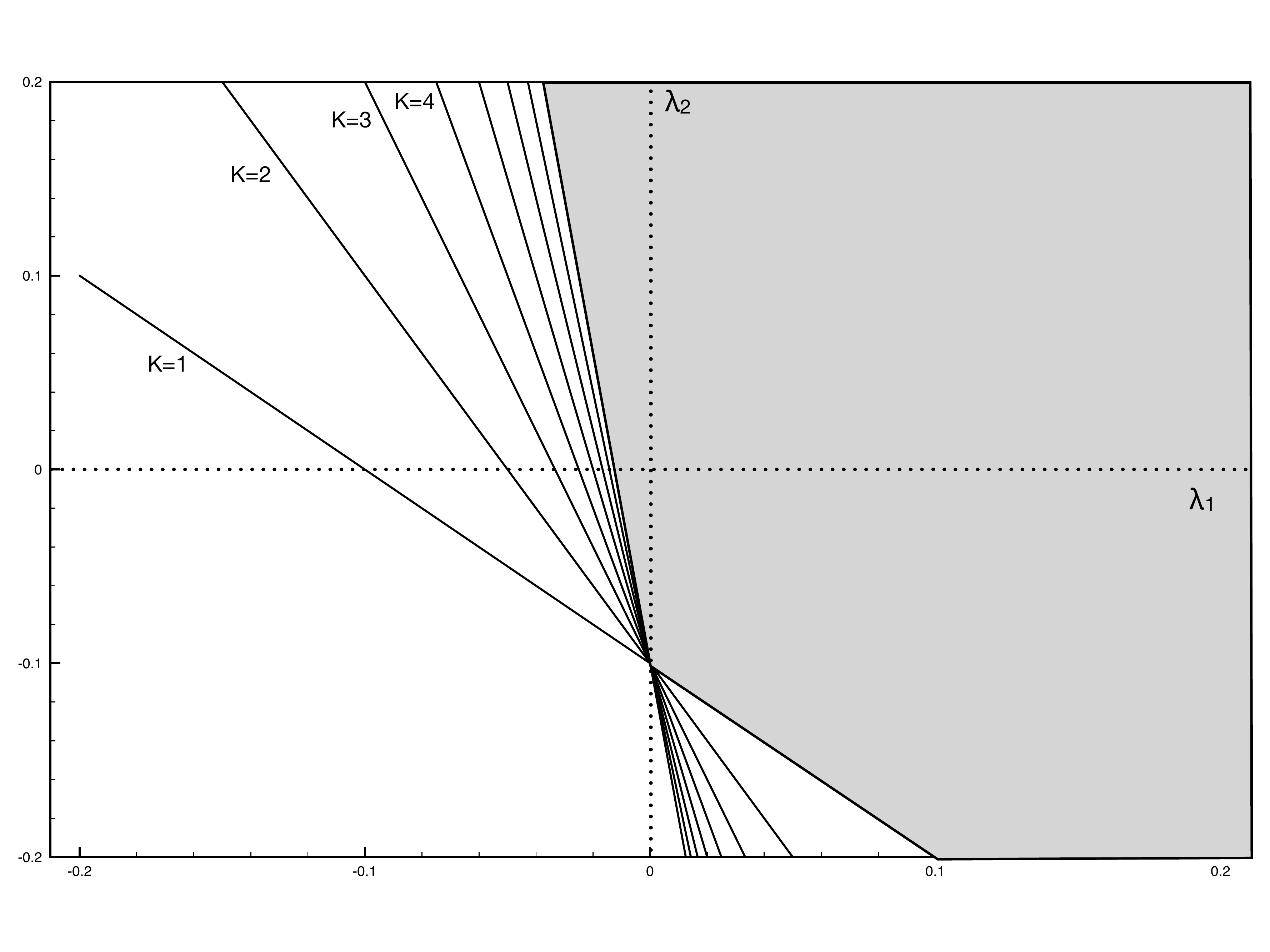}
\caption{\label{StabLines} Stability lines for $K \leq 4$. The grey area is the stable region. }
\end{center}
\end{figure}

This is illustrated in Fig. \ref{StabLines} for an example  with a maximal $K$ of 8. Here the $\lambda_1/\lambda_2-\hbox{plane}$ is shown, and $Q=\rho_2\lambda_2+\Delta$. The parameter $\Delta$ is controlled
by the remaining coupling constants, and in the plot we have chosen $\Delta$ positive. The lines intersect the $\lambda_2$ axis at $\lambda_2=-\Delta/\rho_2$. The grey zone is the region where the 
potential is bounded, for fixed $\lambda_2, \lambda_4$ and $\lambda_5$.
Moving to the right along horizontal lines below the common intersection point one first encounters unbounded territory for $K=1$; above that point it happens for maximal $K$. This nicely illustrates
how  either the maximal or the minimal $K$ solution dominates.
A classical solution defines a boundary line between bounded and unbounded regions if just to the right of that line the solution is a global minimum. Since all solutions with $K=1$ or $K=K_{\rm max}$
are the global minimum for suitable values of the couplings, all such solutions mark the boundary between bounded and unbounded somewhere in coupling space (as before, with the exception of ${\rm C}_y$).

The converse is also true. If one moves in coupling space from a bounded region to an unbounded region, there should be 
classical solution corresponding to the boundary that separates the two regions. To see this more clearly, consider a situation where on coupling, $\lambda_i$, is just at the edge of stability for a value $\lambda_i=c$. 
Then if $\lambda_i$ is made slightly smaller still,
$\lambda_i=c-\eps$,  there is a direction in field space that is unbounded. We can consider a line trough field space in that field direction: a set of fields $\phi_{i\alpha}=t\xi_{i\alpha}$,
so that the potential goes to $-\infty$ for $t\to\infty$. The potential along this direction is
\beqx
V(t)=-\mu^2 t^2 \rho_{\mu} + \tfrac12 \sum_i \lambda_i \rho_i t^4
\eeqx
where $\rho_{\mu}$ and $\rho_i$ are some fixed numbers derived by plugging $\xi_{i\alpha}$ in the various terms in the potential (we use the same notation here as for the parameters $\rho_i$ characterizing
solutions the to equations of motion, because this is just the generalization to general field values).
As a function of $t$
this is a standard quartic potential, which we can analyze as a function of the coupling $\lambda=\sum_i\lambda_i\rho_i$. The minimum is at
\beqx
t^2=\frac{\mu^2\rho_{\mu}}{\lambda};  \ \ \ V_{\rm min}= -\tfrac12 \frac{\mu^4\rho^2_{\mu}}{\lambda}
\eeqx
Hence if $\lambda$ approaches zero from the positive direction, the field goes to $\infty$ and the minimum to $-\infty$. On the other side of the stability line,
for $\lambda_i=c+\eps$, the full potential is bounded from below, and its absolute minimum is a solution to the equations of motion. This absolute minimum may not coincide 
with the minimum along the aforementioned line, but it can only be lower than that. Hence it follows that there is a classical solution that becomes singular exactly at the
boundary line.

This makes it immediately clear that if the potential has more terms, there must be more solutions. If we add another term to the potential, the plot acquires an extra dimension, and an additional hyperplane is needed to constrain the new coupling. This is
true for all terms that are positive definite: $V_1, V_2, V_3$ and $V_4$.  The corresponding coupling constants $\lambda_1, \lambda_2, \lambda_3$ and $\lambda_4$ are bounded from below, but
not from above. But this is not true for $V_5$. This term is not positive definite: the value of $\lambda_5 V_5+\lambda_5^* V_5^*$ can be sign-flipped by replacing
$\phi$ by $\sqrt{i}\phi$. Therefore $|\lambda_5|$ must be bounded. Projected on a real plane in coupling space this implies that $\lambda_5$ must be bounded from above and below. This explains the
appearance of {\em two} additional solutions as soon as $\lambda_5$ is involved. 

\begin{figure}[h!]
\begin{center}
\includegraphics[width=9.5cm]{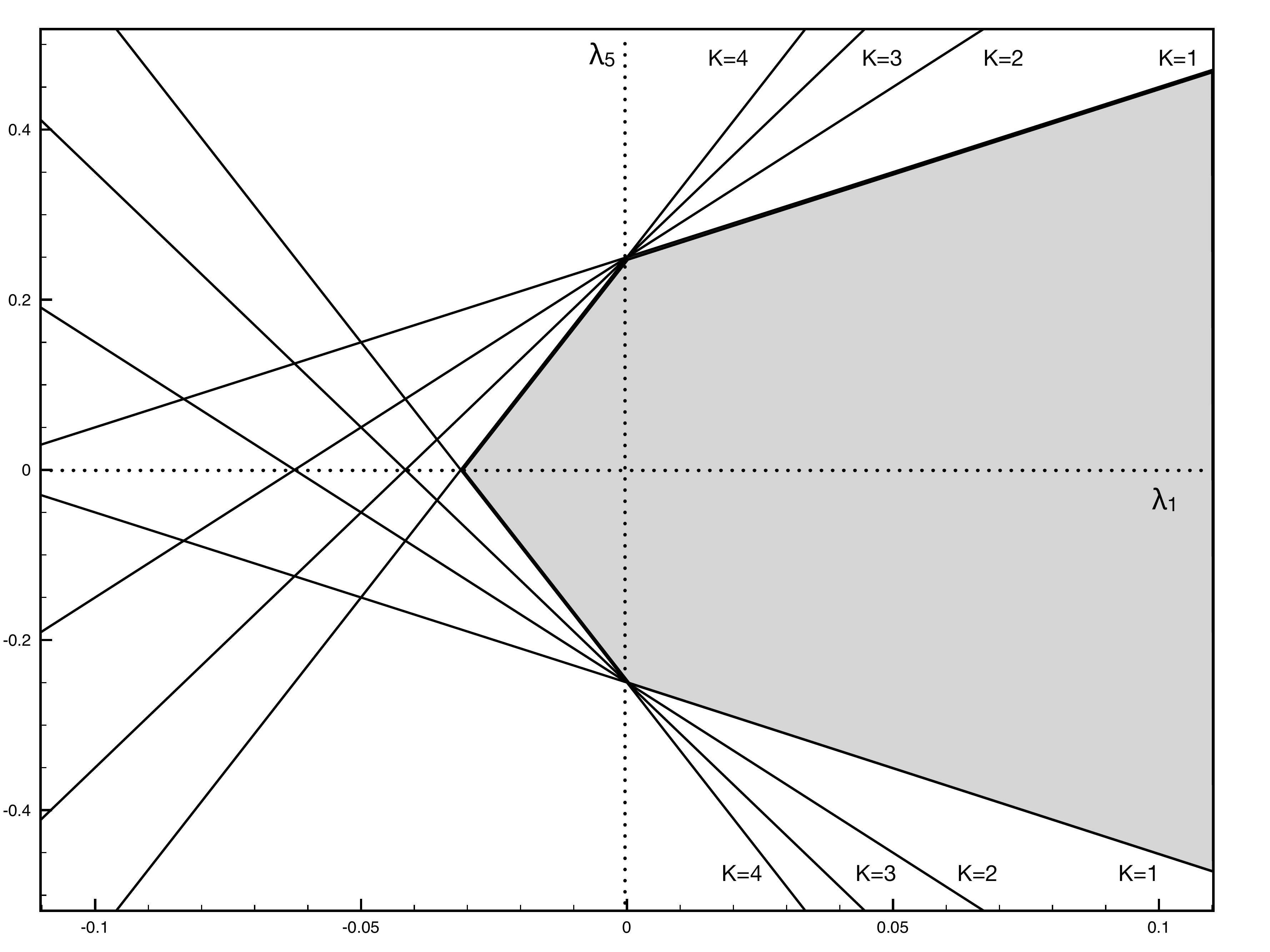}
\caption{\label{StabLinesTwo} Stability lines for $K \leq 4$ in the $\lambda_1,\lambda_5$ plane. The grey area is the stable region. }
\end{center}
\end{figure}

This is shown in figure \ref{StabLinesTwo}, in the plane of $\lambda_1$ and the real part of $\lambda_5$. We have chosen $\lambda_2+\lambda_3+\lambda_4 < 0$, and these parameters are kept fixed. 
Furthermore $K_{\rm max}=4$.
In this situation
the stable region is bounded  by four lines: the first solution with $K=1$, the first solution with $K=K_{\rm max}$, the second solution with $K=K_{\rm max}$ and the second solution with $K=1$. 
The terms ``first" and ``second" solution refer to the two straight lines that exist for every allowed value of $K$ in the real projection. In terms of complex $\lambda_5$, these lines become cones, and the two real solutions are connected by rotations in the complex plane.
The solution with block matrix ${\rm C}_{iy}$ corresponds to the cone opening towards positive $\lambda_1$. This solution provides the boundary of the stability region. The other solution open towards negative $\lambda_1$ and
has the block matrix ${\rm C}_{y}$. It just provides local minima or saddle points.

\subsection{Bi-fundamentals: Summary}\label{BiSum}

In this section we considered Higgs symmetry breaking for gauge groups $G(N) \times H(M)$ with a bi-fundamental Higgs. The previous sections provide answers to the following questions:
\begin{itemize}
\item{What are the stationary points in the potential, and to which subgroup does the group break in those points?}
\item{How do representations decompose under this breaking?}
\item{What is the global minimum?}
\end{itemize}
Because of the length of this section we summarize here how one can obtain this information, without having to read all the arguments.

The answer to the first question is in table \ref{RhoTable}. 
To illustrate how this table is used, we give an example. Suppose one starts with a group $O(N)\times U(M)$. There are two lines in the table with that group
in the first column, which means that there are two classes of minima that can occur in the potential. Suppose we take the second class, with $ \mathbbm{X}=B$. In columns 8 and 9 we find the number of rows $p$ and columns $q$ of  $\mathbbm{X}$. If we take $K$ diagonal copies of $\mathbbm{X}$, there are $N-pK$ rows and $M-qK$ columns of the Higgs v.e.v. matrix that are zero. Hence a subgroup $O(N-pK) \subset O(N)$ and $U(M-qK) \subset U(M)$ remains unbroken. The remainder of these groups is then broken in such a way that the action of $O(pK)$ on the v.e.v. compensates that action of $U(qK)$. This combined action is specified in the column ``Subgroup". The final result, in this particular example, is
\beqx
O(N)\times U(M) \rightarrow O(N-2K)\times U(M-K)\times U(K)
\eeqx

The decomposition of representations $(R_1,R_2)$ of $O(N)\times U(M)$  is obtained as follows. First one decomposes 
$O(N)$ to $O(N-2K) \times O(2K)$ and $U(M)$ to $U(M-K)\times U(K)$. This is just the standard embedding (\ref{BraneSplit}), which can easily be applied to $R_1$ and $R_2$. Next one breaks the remainders of the left and right group to the common subgroup. In this example, that is the breaking
$O(2K) \times U(K)$ to $U(K)$.  The embeddings in the left and right factor are specified in the last column in table \ref{RhoTable}, and the 
numbers in this column refer to the embeddings listed in table \ref{SubgroupTable}. In this example one needs on the left the non-trivial, but well-known embedding of $U(K) \subset O(2K)$, whereas on the right the embedding is trivial, $U(K)\subset U(K)$.  These embeddings must be applied to all the components of $R_1$ and $R_2$, and finally the resulting left and right representations are tensored in $O(N-2K)\times U(M-K)\times U(K)$.

What is the global minimum depends on a fairly complicated way of the relative values of the coupling constants, and we did not attempt to give exact analytical rules for that. It is much easier to determine that numerically using Eqs. (\ref{GeneralEnergy}) and (\ref{Qdef}). But there are two useful general statements. Only stationary points with $K=1$ ot $K$ maximal can be the global minimum.
If there is more than one class of solutions ({\it i.e.} more than one line in table $\ref{RhoTable}$), then each class can occur as the global minimum for
appropriate choices of coupling constants, with the exception of the class with $\mathbbm{X}={\rm C}_y$.

\section{Rank-2 tensors}

In this section we will deal with self-intersecting branes, that give rise to rank-2 tensors. The allowed tensors are dependent on 
the allowed open string endpoints and on the symmetrization. Unitary branes allow two kinds of endpoints that are each other's conjugates, real and symplectic branes have only one. Furthermore one 
can in some cases remove traces to get irreducible representations. The possibilities are listed here
\beqx
\begin{array}{llll}
 U(N)\quad    &\hbox {Hermitean, traceless, complex} &  \phi=\phi^{\dagger},\ \phi_{i}^{~i}=0;  &\phi'=U\phi U^{\dagger} \\
 U(N) \quad     &\hbox {Symmetric, complex}&   \phi=\phi^T;  &\phi'=U\phi U^{T} \\
 U(N) \quad     &\hbox {Anti-symm. complex}&   \phi=-\phi^T;  & \phi'=U\phi U^{T}   \\
 O(N)\quad     & \hbox {Symmetric, traceless, real}&   \phi=\phi^T,\ \phi_{ii}=0; & \phi'=O\phi O^{T} \\
 O(N) \quad    & \hbox {Anti-symm. real}&   \phi=-\phi^T; &\phi'=O\phi O^{T}   \\
 USp(N)\quad    &  \hbox {Symmetric, quaternionic}&   \phi=\phi^T,\ &\phi'=S\phi S^{T} \\
 USp(N)\quad   &   \hbox {Anti-symm. quaternionic, traceless}&   \phi=-\phi^T; \phi_{ij}\Omega_{ij}=0; &\phi'=S\phi S^{T}   
 \end{array}
\eeqx
The matrices $U$, $O$ and $S$ are unitary, orthogonal (unitary and real) and symplectic (unitary and quaternionic) respectively. The matrix $\Omega$ is
the symplectic metric, defined in section \ref{InvCon}. 

\subsection{(Skew)-diagonalization}\label{Diagonal}

The last column shows the gauge transformation of the Higgs field $\phi$. A very useful fact is that in {\it all} these cases the matrix $\phi$ can be diagonalized 
or skew-diagonalized by these transformations. This is well-known for Hermitean matrices and unitary transformations, and for real symmetric matrices and orthogonal transformations. In both
cases one gets a real diagonal matrix, with diagonal elements of either sign. Also well-known is the fact that anti-symmetric real matrices can be brought in skew-diagonal form using orthogonal
transformations. This means that they consist of a diagonal of $2\times 2$ blocks of the form 
\beq\label{Skew}
a_i 
\begin{pmatrix}
0 & 1 \\ 
-1 & 0
\end{pmatrix}
\eeq
where $\lambda_i >0$ plus a number of vanishing $1\times 1$ blocks.  For quaternionic matrices subject to symplectic transformations essentially the same results hold as for
real matrices and orthogonal transformations. If they are symmetric they can be diagonalized in terms of $2\times 2$ blocks of the form \cite{Zhang}
\beq\label{SkewDiag}
a_i 
\begin{pmatrix}
1 & 0 \\ 
0 & 1
\end{pmatrix}
\eeq
with $a_i$ real, and if they are anti-symmetric they can be skew-diagonalized \cite{StanderWiegmann} in precisely the same form as the real anti-symmetric matrices, using blocks (\ref{Skew}). Finally, the result for symmetric complex matrices is
somewhat less well-known, and is called Autonne-Takagi factorization \cite{Autonne,Takagi}. It is used in particle physics to deal with a  Majorana mass matrix.
This was not used in \cite{Li:1973mq}, even though these papers date back to the first quarter of last century. Instead,
this was circumvented by considering the hermitean combination $\phi\phi^{\dagger}$, which can be diagonalized in a more conventional way. The corresponding result for anti-symmetric matrices was proved in 
\cite{StanderWiegmann}. It appears that some of these results have been re-discovered independently, and there may exist earlier references than the ones given here.

\subsection{The potential}

Much of the discussion here is similar to that for the  U-U ,  O-O  and  S-S  bi-fundamentals. The requirement of having just a single field, with a single mass-term, forces us to consider real field for $O(N)$ and
quaternionic fields for $USp(N)$, as was already assumed above. The quartic terms in the potential are precisely the same as in Eqn. (\ref{ThePotential}), with only the $\lambda_1$ and $\lambda_2$ terms. All remaining
ones can be expressed in terms of the $\lambda_2$ term using the reality conditions. The only novelty is that in some cases there is a  cubic term for Hermitean tensors of $U(N)$, symmetric tensors of $O(N)$ and
anti-symmetric tensors of $USp(N)$. This happens for precisely the same fields that can have non-trivial traces.

A quick way of determining all terms is as follows. Bi-fundamentals from different branes give rise to scalar fields $\phi_{i\alpha}$ with two distinct indices. Hence an invariant field combination, such as
those appearing in the potential, has a matrix form consisting of a string of matrices where $\phi$ alternates with its transpose, which may either be $\phi^T$ or $\phi^{\dagger}$. Every invariant must have one or more closed index loops. 
It follows that for bi-fundamentals every closed  index loop must  consist of an even number of matrices $\phi$.  

Rank-2 tensors allow novel contractions between the two indices of $\phi$, because they now belong to the same group. But in addition such tensors always have a definite symmetry under transposition or Hermitean conjugation. 
The new options for index contractions may lead to new 
invariants, but every even index loop can always be brought to a form with alternating fields $\phi$ and $\phi^T$ or $\phi^{\dagger}$, and hence it is necessarily of a form we have already encountered for bi-fundamentals.

Therefore the only possible new terms must involve odd combinations of fields forming closed index loops. Denote such a combination as $(n)$. At second order terms of the form $(2)$    and $(1)(1)$ are possible,
but since we wish to have only one massive propagating field  we must set the combination (1), the trace, equal to zero. Then at third order one can only have $(3)$, and at fourth order there are no new terms at all (the first new
term of even total order is $(3)(3)$).

A cubic term has the general form
\beqx
\kappa \phi_{i}^{~j}\phi_{j}^{~k}\phi_{k}^{~i}
\eeqx 
where the index is raised by a Kronecker $\delta$ or by $\Omega$ for the orthogonal and symplectic cases respectively. In the Hermitean case  the raised index distinguishes the action of $U$ from the action of $U^*$.
Cubic terms do not exist for (anti)-symmetric tensor of $U(N)$, because the index loop cannot be closed in an invariant way.  They vanish 
for anti-symmetric tensors of $O(N)$ and for symmetric symplectic tensors,  because combining the field symmetries and the metric symmetries they are found to be equal to minus themselves.

If their is no cubic term nor a tracelessness condition the discussion is similar to the one for bi-fundamentals. These two complicating factor are closely related: a trace may be thought of as a first-order interaction, and
exists precisely when a cubic interaction exists. Indeed, a non-trivial trace can be dealt with by adding a linear term to the potential as a Lagrange multiplier, as was done in \cite{Li:1973mq}, and by treating its coupling as
a degree of freedom that is varied. 

\subsubsection{Cases without odd invariants}

Without these complications, the entire discussion  in section \ref{Disjoint} applies, and we can view the solution as built out of the basic building blocks $\mathbbm{X}=\mathbbm{1}$  for the symmetric cases and $\mathbbm{X}=i\sigma_2$
for the anti-symmetric ones. Here $\mathbbm{1}$ is $1\times 1$ for $U(N)$ and $2\times 2$ for $USp(N)$
The equations of motion determining the eigenvalues $a_i$ are quadratic and identical for all $i$, but they only determine $a_i$ up to a sign. 

It turns out that  these signs can be rotated away in all cases. For the (anti)-symmetric unitary Higgses this is true because one can choose 
$U_{kl}=\delta_{kl} u_k$, with $u_k=i$ if $a_k$ is a negative eigenvalue. 
This is indeed precisely how one make Majorana masses positive in the Lagrangian. It works only in $U(N)$, not in $SU(N)$.
The sign of a  symmetric $USp(2N)$ block $a_i \mathbbm{1}$ can be 
flipped by means of the $SU(2)$ transformation $i\sigma_3$.  The sign of a matrix $\mathbbm{X}=i\sigma_2$ can be flipped by $O(2)$ rotations $\sigma_3$ or $\sigma_1$ (note that this requires $O(2)$, and that it
does not work in $SO(2)$).  

Since sign flips can be transformed away, this means that there is only one possible non-vanishing eigenvalue. Hence the most  general solution consists of $K$ copies of that block. The energy of this
solution is given by Eqn. (\ref{GeneralEnergy}) with $P=1$ for $U(N)$ symmetric tensors, and $P=2$ in the other three cases. The parameter $\rho_2$ is equal to 1 in all cases (if the block matrix 
is $i\sigma_2$ the special form used in section \ref{InhomEq} cannot be obtained, but one can compute $\rho_2$ explicitly.) Then we get
\begin{eqnarray}\label{SymNoOdd}
V&=&-\frac{K\mu^4}{2(K\lambda_1+\lambda_2)} \ \hbox{for}\ U(N)\ \hbox{(symmetric)} \NO \\
V&=&-\frac{K\mu^4}{(2K\lambda_1+\lambda_2)} \ \hbox{for}\  USp(N)\ \hbox{(sym.)}; \ U(N), O(N)\  \hbox{(anti-sym.)}
\end{eqnarray}
This agrees with \cite{Li:1973mq} when comparable. 

As before the unbroken  symmetry groups fall into two classes: $K$ must be maximal if $\lambda_2 > 0$, and minimal for $\lambda_2 < 0$. The value of $\lambda_1$ is only relevant for stability of the potential: the
denominators in (\ref{SymNoOdd}) should always be positive.  Although only the extreme cases, $K=1$ and $K$ maximal, can occur as absolute minima, all other values of $K$ are extrema (which may be local minima or
saddle points).  It is simplest to list the unbroken groups for all $K$:
\begin{eqnarray}\label{NoOddTraces}
U(N) &\rightarrow&  O(K) \times U(N-K) \quad\quad\quad\quad\quad\ \! \hbox{Symmetric tensor}\nonumber \\
U(N) &\rightarrow& USp(2K)\times U(N-2K)\quad\quad\quad \hbox{Anti-symmetric tensor}\nonumber\\
O(N) &\rightarrow& U(K)\times O(N-2K) \quad\quad\quad\quad\ \  \hbox{Anti-symmetric tensor}\\
USp(N) &\rightarrow& U(K)\times USp(N-2K) \quad\quad\quad\ \ \hbox{Symmetric tensor}\NO
 \end{eqnarray}
 The first two  simply follow from the definition of orthogonal and symplectic groups as the invariance groups of a metric $h$, $UhU^T=h$, where $h$ is either the unit matrix 
 or the anti-symmetric unit matrix $\mathbbm{1} \otimes i\sigma_2$ (see appendix). In the last two cases the subgroups is the simultaneous unitary invariance group of 
 both a symmetric matrix $\mathbbm{1}\otimes \sigma_1$ and the anti-symmetric matrix $\mathbbm{1}\otimes i\sigma_2$. One of these matrices defines the original unbroken gauge group,
 and the other is the Higgs v.e.v.
 
 Writing the subgroups for all $K$ clarifies some special features of the special case $K=1$, especially regarding the global group. In particular: the symmetric tensor breaks $U(N)$ to $U(N-1)$ times a $\mathbbm{Z}_2$ symmetry,
 the $U(N)$ matrix ${\rm diag}(-1,1,\ldots,1)$. This is $O(K)$ for $K=1$. The anti-symmetric tensor breaks the first two components of $U(N)$ to $USp(2)$.  Without the result for all $K$ one might have called this $SU(2)$, which
 is correct, but gives the incorrect impression that special unitary groups appear after symmetry breaking. However, this happens only for the special unitary group $SU(2)$, which must be interpreted as $USp(2)$. Apart from this
 isomorphism, we never get $SU(K)$ factors in the unbroken group. Finally in the third case the first factor is $U(1)$ and not $SO(2)$. These are isomorphic as groups, but using the correct notation avoids some subtle mistakes.
 First of all we see immediately that the first factor is $SO(2)$ and not $O(2)$, which is locally isomorphic to, but globally
 different from $SO(2)$.  Secondly, we never get special  orthogonal groups, except $SO(2)$, because of the isomorphism with $U(1)$. This may not seem important, but the implication of not getting special unitary or orthogonal
 groups is that all broken subgroups can be realized in terms of membranes.  We will however not explore this point further in this paper.

\subsubsection{Cases with odd invariants}

In this category we have three kinds of Higgs fields, Hermitean $U(N)$ tensors, real symmetric $O(N)$ tensor and anti-symmetric $USp(N)$ tensors.
A detailed analysis of the first two is in appendix B of \cite{Li:1973mq}, and the only novelty here is the anti-symmetric, symplectic case.  However, in all three cases
one ends up with the same equations, and hence the conclusions are also the same. We will illustrate this for the anti-symmetric tensor of $USp(2N)$, the only new case.
The potential is
\beqx
V= -\mu^2 {\rm Tr} \phi\phi^{\dagger}   + \tfrac12  \lambda_1 ({\rm Tr} \phi\phi^{\dagger})^2+   \tfrac23 \kappa{\rm Tr} (\phi\Omega)^3 + \tfrac12  \lambda_2 {\rm Tr} (\phi\phi^{\dagger})^2 - 2g {\rm Tr} \phi\Omega
\eeqx
The last term is a Lagrange multiplier; demanding stability with respect to $g$-variations yield the trace condition. We have included a factor of 2 for
comparison with    \cite{Li:1973mq}, because the rest of the potential also differs by a factor of 2.
Substituting the skew diagonal form (\ref{SkewDiag}) we get
\beqx
V= -2\mu^2\sum_i a_i^2   +  2\lambda_1\left[ \sum_i a_i^2\right]^2 + \tfrac43\kappa \sum_i a_i^3 +\lambda_2  \sum_i a_i^4 - 4g \sum_i a_i
\eeqx
The equations of motion for the eigenvalues $a_i$ are
\beq\label{TraceEOM}
 - \mu^2 a_i   +  2\lambda_1a_i \left[ \sum_j a_i^2\right] +  \kappa a_i^2+ \lambda_2   a_i^3 - g =0
\eeq
This is the same\footnote{with $\kappa=\lambda_3$, and after correcting a typo in (B17)} equation as (B18) of \cite{Li:1973mq} apart from a factor of 2 in front of the second term. This factor  is just the coefficient $P$ introduced in section \ref{InhomEq}, which indeed is 2 for 
the basic $USp(2N)$ anti-symmetric block, and 1 for the other two cases. 
Hence all results of  \cite{Li:1973mq} go though after a replacement of $\lambda_1$ with $2\lambda_1$.
This implies that two distinct  absolute minima can exist, depending on $\lambda_1,\lambda_2$ and $\kappa$.

One may summarize all three cases at once: the rules and the dynamics are the same for Hermitean fields in $G(N)=U(N)$, symmetric fields in $G(N)=O(N)$ and anti-symmetric fields in 
$G(N)=USp(2N)$. In general, the group $G(N)$  splits into several components using the brane separation embedding (\ref{BraneSplit}). With some  string theory intuition, this means that the stack of $N$ branes  is split into several smaller stacks.
There are stationary points where the group $G(N)$ splits into three parts, but for the global minima there are just two possible unbroken subgroups 
\begin{equation}\label{OddTracesOne}
G(N) \rightarrow G(N\!-\!1)\times G(1)
\end{equation}
and
\begin{eqnarray} \label{OddTracesTwo}
G(N) &\rightarrow& G(\tfrac12 N)\times G(\tfrac12 N) \quad \quad \quad \quad \quad\ (N\ \hbox{even})\NO\\
G(N) &\rightarrow& G(\tfrac12 (N\!+\!1))\times G(\tfrac12 (N\!-\!1)) \quad (N\ \hbox{odd})
\end{eqnarray} 
According to \cite{Li:1973mq}, if $\kappa=0$ and $\lambda_2 > 0$ the second minimum is the lowest one, and when $\kappa$ is increased the first becomes the lowest one. If $\lambda_2 < 0$ we get the first 
minimum, and this remains  true even if $\kappa$ is varied.  Here $\lambda_1$ is assumed to be positive. 

Note that in contrast to all other cases, the non-trivial part of the solution of the equations motion is not a combination of $K$ identical block matrices, but of two (or three, if we also count solutions
that are not absolute minima) distinct $1\times 1$ blocks. The reason why this happens can be understood by considering the weighted  difference of two equations (\ref{TraceEOM}), with variables $a_1$ and $a_2$.
Multiplying the first with $a_2$, the second with $a_1$, and subtracting them we get 
\beqx
\kappa a_1 a_2 (a_1-a_2) + \lambda_2  a_1a_2( a_1-a_2) ( a_1+a_2)+ g(a_1-a_2)=0
\eeqx
In the absence of a cubic term and the Lagrange multiplier term, $\kappa=g=0$, this equation implies that the two eigenvalues must be the same or opposite,
or one of them must vanish. Since in all relevant cases signs can be flipped, it follows that there can exist just one distinct non-vanishing eigenvalue. It is clear that either 
the existence of a cubic term or the tracelessness condition make that argument invalid.

\subsection{Rank-2 tensors: Summary}\label{TwoSum}

The main results of this section are given in Eqns. (\ref{NoOddTraces}), (\ref{OddTracesOne}) and (\ref{OddTracesTwo}). The embeddings used here
are the same ones we already encountered for bi-fundamentals. In (\ref{NoOddTraces}) one first applies the ``brane separation" embedding (\ref{BraneSplit}) to split off the second factor on the right hand side. Then the first factor is broken according to embedding 1,2,3 and 4 of table \ref{SubgroupTable} for the four cases listed in Eqn. (\ref{NoOddTraces}) respectively. The embeddings in (\ref{OddTracesOne}) and (\ref{OddTracesTwo}) are just brane separation embeddings (\ref{BraneSplit}); they are all of the form $G(N)  \rightarrow G(N-K) \times G(K)$. All solutions come with an integer label $K$, but the global minimum only occurs for either $K=1$ or the maximal value of $K$ (depending on the coupling constants, as explained above).  In the cases without odd invariants, the maximal value of $K$ is the one for which the second group factor in (\ref{NoOddTraces}) is minimal or trivial.
In the three cases with odd invariants, with $G(N)  \rightarrow G(N-K) \times G(K)$, what we mean by ``maximal" is the value where $G$ is maximally split, namely
$K=\lfloor N/2\rfloor$ (note that the cases $K=k$ and $K=N-k$ are identical).  We will refer to this as $K={\rm ``max"}$ in all cases.

In order to clarify the comparison with the results of  \cite{Li:1973mq}  we have combined all the results for rank-2 tensors in table \ref{RankTwoTable}, analogous to table III of  \cite{Li:1973mq}, but with rows and columns interchanged. The main differences with \cite{Li:1973mq} are that we start with  $U(N)$ as the unbroken group instead of $SU(N)$, and that we  have included  the results for
symplectic groups.  Furthermore we have left $O(1)$ factors that automatically appear if $K=1$ or $K={\rm ``max"}$ is substituted in the general formula.  They give rise to a $\mathbb{Z}_2$ discrete symmetry.

\begin{table}[!h]
\begin{center}
\begin{tabular}{|l|c|l|l|l|} \hline
Group & $K$ &Symmetric tensor &      Anti-sym.  tensor& Adjoint  \\ \hline
$U(N)$ & 1 & {$U(N\!-\!1)\times O(1)$}  &  $U(N\!-\!2)\times USp(2)$    &    $U(N\!-\!1)\times U(1)$   \\  
$U(2\ell)$ & max &$O(2\ell)$ &  $USp(2\ell)$    &      $U(\ell)\times U(\ell)$  \\                    
$U(2\ell\!+\!1)$ & max & $O(2\ell+1)$ &  $USp(2\ell)\times U(1)$    &     $U(\ell)\times U(\ell+1)$\\
$O(N)$ & 1& $O(N-1)\times O(1)$  &   $O(N-2)\times U(1)$   &   -    \\  
$O(2\ell)$ & max & $O(\ell) \times O(\ell)$&   $U(\ell)$   &    -   \\  
$O(2\ell\!+\!1)$ & max & $O(\ell) \times O(\ell+1)$&   $U(\ell)\times O(1)$    &    -   \\  
$USp(2N)$& 1 & $USp(2N-2)\times U(1)$ &  $USp(2N-2) \times USp(2)$   &   -    \\  
$USp(4\ell)$& max & $U(2\ell)$  &  $USp(2\ell)\times USp(2\ell)$  &   -    \\  
$USp(4\ell\!+\!2)$ & max& $U(2\ell+1)$ &  $USp(2\ell)\times USp(2\ell\!+\!2)$    &  -    \\  
     \hline
\end{tabular}
\caption{Rank-2 tensor breaking patterns.}\label{RankTwoTable}
\end{center}
\end{table}

For $K={\rm ``max"}$ one sometimes has to distinguish even and odd $N$, so we have used two separate lines in the table. Column 2 specifies $K$,
and columns 3,4 and 5 display the results for the  various tensor representations of the Higgs field. Note that the adjoint representations of $O(N)$ and $USp(N)$ are anti-symmetric and symmetric tensors respectively.

There are a  few known errors in table III of \cite{Li:1973mq}. 
In \cite{Elias:1975yd} it was pointed out that a factor $USp(2)$ was overlooked in the antisymmetric tensor breaking of $U(N)$ for $K=1$. Furthermore a $U(1)$ was overlooked in the adjoint breaking of $U(N)$ for $K=1$.  There is one more error not mentioned in \cite{Elias:1975yd}: The anti-symmetric tensor breaking of $U(N)$ for $K={\rm ``max"}$ yields $USp(2\ell)$ and $USp(2\ell)\times U(1)$ for $N$ odd, and not $O(2\ell+1)$ as stated in table III of \cite{Li:1973mq}. This is evidently just a transcription error in table III, because in section IIIc the correct result was given:  $U(2\ell) \to Sp(2\ell)$ for $N=2\ell$ and $SU(2\ell+1) \to Sp(2\ell)$ for $N=2\ell+1$, in agreement with our result (note the use of $SU(2\ell+1)$ in \cite{Li:1973mq} instead of $U(2\ell+1)$ in our case).

\section{Conclusions}

The classic work of \cite{Li:1973mq} from 1973 turns out to have an elegant generalization to all Higgs representations one can ever encounter in intersecting brane models. In all cases without trace conditions
and cubic terms the solutions to the equations of motion are characterized by an integer $K$. The global minimum  of the potential has $K$ either equal to 1 or the maximal value that can be realized. If the two intersecting
brane groups are of different types, there are additional terms in the potential, and for each additional term there is an additional class of solutions. Each class is characterized by it own integer $K$. For suitable parameter
values, each class can provide the absolute minimum, for either maximal or minimal $K$.

The foregoing holds for bi-fundamentals as well as self-intersections. However, in the latter case there are three cases with non-trivial traces and cubic terms, an orthogonal, a unitary and a symplectic one. 
As already shown in \cite{Li:1973mq} there are now more possibilities for extrema of the potential. They are not characterized by a single integer, but by two integers. 
In these extrema, the original group is split into various parts of the same type ({\it i.e.} products of $U(n_i), O(n_i)$ or $USp(n_i)$ if the original group is $U(K), O(K)$ or $USp(2K)$). 
The maximal number of parts one encounters is three, but in the absolute minimum the group is split into two parts only. 
Depending on the coupling constant values, it is either split in to two equal (if $K$ is even) or almost equal parts (if $K$ is odd), or  it is split in the smallest possible non-trivial part times
the largest possible part.

It turns out that {\em in all cases} the unbroken subgroup can be written as a product of $U(K)$, $O(K)$ and $USp(2K)$ factors. In particular, there are no special unitary or special orthogonal groups, except as a result
of certain low-rank isomorphisms. 
This is likely to have a nice interpretation in turns of brane dynamics and the
phenomenon known as ``brane recombination", but we leave further exploration of this point to future work, since here we only intended to address purely field-theoretic issues.

\vskip 2.truecm
\noindent
{\bf Acknowledgements:}
It is a pleasure to thank Beatriz Gato-Rivera for discussions and contributions during an early stage of this work, and IFF-CSIC Madrid, where part of this work was done, for hospitality. 
\vskip .2in
\noindent

\appendix

\section{Orthogonal and Symplectic groups}

Here we collect some facts about symplectic groups, and some related features of orthogonal groups acting on spaces of even dimension.
Consider the subset of unitary $2K\times 2K$ matrices, $U(2K)$,  that satisfies the following restriction.
\begin{equation}\label{GroupRestriction}
U h U^T = h
\end{equation}
For any $h$ this defines a subgroup of $U(2K)$. Standard choices are $h_S=\mathbbm{1}$, in which case the subgroup is $O(2K)$, and the matrix
\begin{equation}\label{hAdef}
h_A=
\begin{pmatrix}
0 & \mathbbm{1}\\
\mathbbm{-1} & 0
\end{pmatrix} \ .
\end{equation}
The resulting subgroup is called $USp(2K)$. In the definition of the orthogonal groups the restriction to even dimensions is not necessary, but the special
features of interest here only hold in even dimensions. 

To put the two groups on similar footing we may choose a different basis in the orthogonal case.
In even dimensions we may choose instead of the metric $h_S=\mathbbm{1}$ the symmetric matrix
\beqx
\tilde h_S=
\begin{pmatrix}
0 & \mathbbm{1}\\
\mathbbm{1} & 0
\end{pmatrix} \ .
\eeqx
This defines a different subgroup of $U(2N)$ which is isomorphic to $O(2N)$. Their elements $\tilde U$ are related by a unitary matrix
$Z$ in the following: $\tilde U =Z^{\dagger} U Z$. If $UhU^T=h$, then $\tilde U\tilde hU^T=\tilde h$, with $\tilde h= Z^{\dagger} h Z^*$.
If one chooses
A useful choice is the matrix 
\beq\label{XTransform}
Z=\tfrac{1}{\sqrt 2}\begin{pmatrix}
 \mathbbm{1} &  \mathbbm{1}\\
 i \mathbbm{1} & -i\mathbbm{1}
\end{pmatrix}
\eeq
then $h_S$ is transformed to $\tilde h_S$.

\subsection{The Lie-algebra}

We work out the Lie algebra for a metric
\beq\label{Simultaneous}
h=
\begin{pmatrix}
0 & \mathbbm{1}\\
\epsilon\mathbbm{1} & 0
\end{pmatrix} \ 
\eeq
A generator $T$ of the Lie algebra is found to have the following form.
\begin{equation}
T=
\begin{pmatrix} H & S^{\dagger} \\
S& -H^{*}   
\end{pmatrix}
\end{equation}
where $H$ is hermitean and $S$ is complex and satisfies 
\beqx
S^T = -\epsilon S
\eeqx
The total number of real parameters of this Lie algebra is \beqx K^2+2\times \tfrac12 K(K-\epsilon)=\tfrac12 (2K)(2K-\epsilon) \eeqx
which is indeed the correct answer for $O(2K)$ ($\epsilon=1)$ and $USp(2K)$ ($\epsilon =-1)$.

The subset of generators with $S=0$ generate a $U(K)$ sub-algebra, with corresponding group matrices
\beqx
U=\begin{pmatrix}
Y & 0 \\
0 & Y^*
\end{pmatrix}
\eeqx
where $Y$ is unitary.

\subsection{Reality conditions}
Consider now any vector $\phi$ that $U$ acts on.
If $U \in {\cal H}$ we can consistently limit the space on which it acts, in the following way
\begin{equation}
\phi=h\phi^*
\end{equation}
Because 
\begin{equation}
U\phi=Uh\phi^*=hU^{T-1}\phi^*=hU^*\phi^*=h(U\phi)^*
\end{equation}
Hence the action of ${\cal H}$ preserves the condition $\phi=h\phi^*$.
But we must also satisfy the consistency condition
\begin{equation}
\phi=h\phi^*=h(h\phi^*)^*=hh^*\phi
\end{equation}
Since this must hold for arbitrary vectors, this implies that $hh^*={\mathbbm{1}}$. A symplectic (anti-symmetric) metric does not satisfy this.
Note that $h h^*=h (h^{\dagger})^T$. If $h$ is anti-symmetric, we get $-hh^{\dagger}$, which is negative definite. Hence we cannot impose
such a condition on  vectors. For an orthogonal group embedded with $h_S=\mathbbm{1}$ this restricts $\phi$ to be real. In we use the symmetric metric
$\tilde h_S$, the reality condition implies that $\phi$, written as a row vector, has the form $(\chi,\chi^*)$ where $\chi$ is a $K$-dimensional complex vector.

Although one cannot impose reality conditions on symplectic vectors, one can impose them on rank-2 tensors. Suppose
a tensor $T$ transforms as $T\rightarrow UTV^T$, where $U$ and $V$ are symplectic matrices, not necessarily elements of the same group,
and not necessarily of equal dimension. Suppose they satisfy $Uh_1U^T=h_1$ and $Vh_2V^T=h_2$. Now we can impose the condition
\beq\label{SympReal}
T=h_1T^* h_2^T
\eeq
It is easy to check that this condition
is preserved by the transformation. This condition can be imposed consistently if both matrices $h_1$ and $h_2$ are symmetric, or if both are anti-symmetric. 
The former is simply the standard reality condition for orthogonal group tensor representations (if one uses $h_1=\mathbbm{1}$ $h_2=\mathbbm{1}$). We will refer to the latter as a symplectic reality condition. 

In some cases a basis exists where the symplectic reality condition become an ordinary reality condition.
For example rank-2 tensor combinations of symplectic vectors are real representations, which means that such a
basis does indeed exist. Similarly, the representation $(V,V)$ in $USp(2)\times USp(2)$ is real; it is the vector representation
of $SO(4)$. This fact may or may not extend to $USp(2N)\times USp(2M)$, but in any case the real basis is not useful for 
our purposes, because it combines transformations of the two group factors.

\subsection{Quaternionic  basis}\label{QuatBas}

For both orthogonal and symplectic groups and algebras, there is another useful basis. It is obtained by reordering the original basis as $(1,K+1,2,K+2,\ldots,K-1,2K)$. This splits
the Lie-algebra matrices into $2\times 2$ blocks. 
The matrices $h_S$ and $h_A$ take the form $\mathbbm{1}_K \otimes \sigma_1$ and $\mathbbm{1}_K \otimes i\sigma_2$ respectively. 
The block-diagonal symplectic transformations form a subgroup $SU(2)^K$, which is isomorphic to $USp(2)^K$.  The block-diagonal orthogonal transformations form a subgroup $U(1)^K$ or,
equivalently, $SO(2)^K$.

We will focus on the symplectic case.
 In the new basis the Lie algebra generators
can be written in terms of $2\times 2$ blocks of the form
\begin{equation*}
\begin{pmatrix}
a & b^* \\
b & -a^*
\end{pmatrix}
\end{equation*}
Hence they have 4 parameters each, except on the diagonal, where $a$ must be real.  

The elements of the block matrix can be interpreted as quaternions.
Quaternions are numbers of the form 
\beqx
a+ib+jc+kd
\eeqx
with $a,b,c,d \in \mathbbm{R}$ and with $i^2=j^2=k^2=ijk=-1$.
They can be represented by $2\times 2$ matrices
\beqa
\begin{pmatrix}
a+bi & c+di \\
-c+di & a-bi
\end{pmatrix}
\eeqa
This means that the multiplication of any two such matrices yields the answer representing the product of the corresponding quaternions.
Note that the Lie algebra generators are not precisely quaternions, but ``imaginary" quaternions, quaternions multiplied by $i$. But the group
elements are quaternions. This is analogous with orthogonal groups in a real basis: the group elements are real, and the Lie algebra matrices
are purely imaginary (in the standard physics convention).

\def\Qp{{\rm Q}_{+}}
\def\Qm{{\rm Q}_{-}}
%
%

Now consider a rank-2 tensor $T$ satisfying the symplectic reality condition (\ref{SympReal}). We can write both $h_1$ and $h_2$ in block-diagonal
form. They are then equal to $h_1=\mathbbm{1}_M \otimes (i \sigma_2)$ and $h_2=\mathbbm{1}_N \otimes (i \sigma_2)$, where $M$ may be different from $N$.
It is now easy to see that (\ref{SympReal}) implies that the $2\times 2$ blocks of $T$ are quaternions. Instead of (\ref{SympReal}) one may define a reality
condition with an opposite sign. If $T$ satisfies that condition, it is built out of quaternions times $i$. This is the analog of splitting a complex field in real and imaginary parts.

 \subsection{Special forms of vectors}

 It is well-know that a real vector can be rotated into any direction using orthogonal rotations. This implies that any real vector can be orthogonally rotated to the form
\beq\label{SpecialVector}
(r,0,\ldots,0)
\eeq
where $r$ is real and positive. The same result holds for $U(N)$ acting on complex vectors. There is also an analogous result for symplectic transformations acting on quaternions.
This can be seen as follows. Consider a quaternionic vector $(q_1,\ldots,q_K)$, where $q_i$ are quaternions. Up to normalization, quaternions are $SU(2)$ group elements. Therefore, using the
$USp(2K)$ subgroup $SU(2)^K$ we can rotate all $q_i$ to the form $(d_1\mathbbm{1},\ldots,d_K\mathbbm{1})$, with $d_i \in \mathbb{R}$. Now we use the $U(K)$ subgroup of $USp(2K)$, or in fact
just its $O(K)$ subgroup. This acts on $(d_i,\ldots,d_k)$ as a vector, and it can therefore rotate this vector so that only the first component is non-zero. Hence we get (\ref{SpecialVector}) with $r$ interpreted as a 
$r$ times a unit quaternion.

\subsubsection{Orthogonal and symplectic transformations of complex vectors}

In addition to this we also need such results for orthogonal and symplectic transformations acting on {\em complex} vectors. 
With orthogonal transformations acting on complex vectors we can first rotate all imaginary parts into the first entry, and then using $O(N-1)$ rotations 
on the last $N-1$ components rotate the remaining real components into the second entry. Then the simplified form is
\beq\label{RealSpecialVector}
(x,r,0,\ldots,0)
\eeq
With $x$ complex and $r$ real. Note that $x$ must have an imaginary part; otherwise we can simplify further and rotate $r$ into $x$.

For symplectic transformations one can show that any vector can always be rotated to the form (\ref{SpecialVector}), but this time with $r$ interpreted as a real number, and not as a quaternion. 
This is shown as follows. Using the $SU(2)^K$ subgroup we can rotate in each block the 2-dimensional complex vectors to the form $(r,0)$. This brings a general complex vector to the form
$(r_1,0,r_2,0,\ldots,r_K,0)$. Now, using $U(K)$ transformations, we can rotate away $r_2,\ldots r_K$. These $U(K)$ transformations also act on the odd components, but since they all vanish already
this is irrelevant.
It may seem that with $USp(2K)$ we can do the same as with $U(2K)$, but this is not true.
The difference with unitary rotations only becomes apparent
when one tries to bring a second vector in a simplified form. The generic simplified form of two vectors in $U(2K)$ is
\beqx
\begin{pmatrix}
r_1&0&0&\ldots&0\\
x&r_2&0&\ldots&0\\
\end{pmatrix}
\eeqx
where $x$ is a complex number. For $USp(2K)$ this is
\beqx
\begin{pmatrix}
r_1&0    &0    &0&\ldots&0\\
x_1&x_2&r_2&0&\ldots&0\\
\end{pmatrix}
\eeqx
since the first vector fixes rotations on the first two components, whereas in $U(2K)$ it only fixes the first components.

\
\bibliography{Landscape}{}

\bibliographystyle{unsrt}

\end{document}